\documentclass[12pt, preprint]{aastex}
\usepackage{emulateapj5}
\usepackage{amsmath}
\usepackage{graphicx}
\usepackage{natbib}
\shortauthors{L. -G. Strolger et al.}
\shorttitle{The Hubble Higher-$z$ Supernova Search Project}

\begin{document}
\title{The Hubble Higher-{\it\lowercase{z}} Supernova Search:\\ Supernovae to
{\it\lowercase{z}} $\approx 1.6$ and Constraints on Type I\lowercase{a} Progenitor
Models
\footnote{Based on observations with the NASA/ESA {\it Hubble
Space Telescope}, obtained at the Space Telescope Science Institute, which is
operated by AURA, Inc., under NASA contract NAS 5-26555} 
}

\author{Louis-Gregory~Strolger$^{2}$, Adam~G.~Riess$^{2}$,
Tomas~Dahlen$^{2}$, Mario~Livio$^{2}$, Nino~Panagia$^{2,3}$,
Peter~Challis$^{4}$, John~L.~Tonry$^{5}$, Alexei~V.~Filippenko$^{6}$,
Ryan~Chornock$^{6}$, Henry~Ferguson$^{2}$, Anton~Koekemoer$^{2}$,
Bahram~Mobasher$^{2,3}$, Mark~Dickinson$^{2}$, Mauro~Giavalisco$^{2}$,
Stefano~Casertano$^{2}$, Richard~Hook$^{7}$, Stephane~Blondin$^{8}$,
Bruno~Leibundgut$^{8}$, Mario~Nonino$^{9}$, Piero~Rosati$^{8}$,
Hyron~Spinrad$^{6}$, Charles~C.~Steidel$^{10}$, Daniel~Stern$^{11}$,
Peter~M.~Garnavich$^{12}$, Thomas~Matheson$^{4}$, Norman~Grogin$^{13}$, Ann~Hornschemeier$^{13}$,
Claudia~Kretchmer$^{13}$, Victoria~G.~Laidler$^{14}$, Kyoungsoo~Lee$^{13}$,
Ray~Lucas$^{2}$, Duilia~de~Mello$^{13}$, Leonidas~A.~Moustakas$^{2}$,
Swara~Ravindranath$^{2}$, Marin~Richardson$^{2}$, and~Edward~Taylor$^{15}$}

\affil{$^{2}$Space Telescope Science Institute, 3700 San Martin Drive,
Baltimore, MD 21218 (email: strolger@stsci.edu).}

\affil{$^{3}$Affiliated with the Space Telescope Division of the European
Space Agency, ESTEC, Noordwijk, the Netherlands.}

\affil{$^{4}$Harvard-Smithsonian Center for Astrophysics, 60 Garden Street,
Cambridge, MA 02138.}

\affil{$^{5}$University of Hawaii, Institute for Astronomy, 2680 Woodlawn
Drive, Honolulu, HI 96822.}

\affil{$^{6}$Department of Astronomy, University of California, 601 Campbell
Hall, Berkeley, CA 94720-3411.}

\affil{$^{7}$Space Telescope - European Coordinating Facility, European Southern
Observatory, Karl Schwarzschild Str.-2, D-85748, Garching, Germany.}

\affil{$^{8}$European Southern Observatory, Karl-Schwarzschild-Str. 2, D-85748,
Garching, Germany.}

\affil{$^{9}$INAF, Astronomical Observatory of Trieste, Via Tiepolo 11 34131
Trieste, Italy}

\affil{$^{10}$Department of Astronomy, California Institute of Technology, MS
105-24, Pasadena, CA 91125.}

\affil{$^{11}$Jet Propulsion Laboratory, MS 169-506, California Institute of
Technology, Pasadena, CA 91109.}

\affil{$^{12}$University of Notre Dame, 225 Nieuwland Science Hall, Notre Dame,
IN 46556}

\affil{$^{13}$Johns Hopkins University, Dept. of Physics and Astronomy, 3400
N. Charles Street, Baltimore, MD 21218.}

\affil{$^{14}$Computer Sciences Corporation at Space Telescope Science
Institute, 3700 San Martin Drive, Baltimore, MD 21218.}

\affil{$^{15}$University of Melbourne, School of Physics, Victoria 3010,
Australia.}

\begin{abstract}
We present results from the Hubble Higher-$z$ Supernova Search, the first
space-based open field survey for supernovae (SNe). In cooperation with the
Great Observatories Origins Deep Survey, we have used the {\it Hubble Space
Telescope} with the Advanced Camera for Surveys to cover $\sim 300$ square
arcmin in the area of the Chandra Deep Field South and the Hubble Deep Field
North on five separate search epochs (separated by $\sim45$ day intervals) to a
limiting magnitude of $F850LP \approx 26$. These deep observations have allowed
us to discover 42 SNe in the redshift range $0.2 < z < 1.6$. As these data span
a large range in redshift, they are ideal for testing the validity of Type Ia
supernova progenitor models with the distribution of expected ``delay times,''
from progenitor star formation to SN~Ia explosion, and the SN rates these
models predict. Through a Bayesian maximum likelihood test, we determine which
delay-time models best reproduce the redshift distribution of SNe~Ia discovered
in this survey. We find that models that require a large fraction of ``prompt'' (less than 2 Gyr)
SNe~Ia poorly reproduce the observed redshift distribution and are rejected at
$>95\%$ confidence. We find that Gaussian models best fit the observed data for mean
delay times in the range of  3 to 4 Gyr.

 \end{abstract}


\section{Introduction}

Type Ia supernovae (SNe~Ia) have proven that they are unequivocally suited as
precise distance indicators, ideal for probing the vast distances necessary to
measure the expansion history of the Universe. The results of the High-$z$
Supernova Search Team~\citep{1998AJ....116.1009R} and the Supernova Cosmology
Project~\citep{Perlmutter1999565} have astonishingly shown that the Universe is
not decelerating (and therefore not matter dominated), but is apparently
accelerating, driven apart by a dominant negative pressure, or ``dark
energy.'' Complementary results from the cosmic microwave background by
WMAP~\citep{2003ApJS..148....1B} and large-scale structure from
2dF~\citep{2001Natur.410..169P, 2001MNRAS.327.1297P, 2002MNRAS.330L..29E}
congruously show evidence for a low matter density ($\Omega_{M}=0.3$) and a
non-zero cosmological constant ($\Omega_{\Lambda}=0.7$), but neither directly
require the presence of dark energy.

However, it is possible that there are astrophysical effects which allow SNe~Ia
to appear systematically fainter with distance and therefore mimic the most
convincing evidence for the existence of dark energy. A pervasive screen of
``gray dust'' scattered within the intergalactic medium could make SNe~Ia seem
dim, but show little corresponding
reddening~\citep{1999ApJ...512L..19A}. Alternatively, the progenitor systems of
SNe~Ia could be changing with time, resulting in evolving populations of
events, and possibly necessitating modifications to the empirical correlations
which are currently used to make SNe~Ia precise standard candles. To date, the
investigations of either effect have only provided contrary evidence,
disfavoring popular intergalactic dust models~\citep{2000ApJ...536...62R}, and
statistically showing strong similarity in SN~Ia characteristics at all age
epochs, locally and at $\langle z\rangle \approx
0.5$~\citep{1998AJ....116.1009R,Perlmutter1999565,2000ApJ...536...62R,
Aldering20001192110A,2003MNRAS.340.1057S}, but neither has been conclusively
ruled out.

A simple test of the high-redshift survey results would be to search for SNe~Ia
at even higher redshifts, beyond $z \approx 1$. In the range $1 < z < 2$, we
should observe SNe~Ia exploding in an epoch of cosmic deceleration, thus
becoming relatively brighter than at lower redshifts. This is expected to be
unmistakably distinguishable from simple astrophysical challenges to the SN~Ia
conclusion. Indeed, results from 19 SNe~Ia observed in the range $0.7 < z <
1.2$ from the latest High-$z$ Supernova survey~\citep{2003ApJ...594....1T} and
in the IfA survey~\citep{Barris:2003dq} show indications of past deceleration,
but these SNe represent the highest-redshift bin attainable from the ground, in
which confident identification and light-curve parameters are pushed to their
limits. To thoroughly and reliably survey SNe~Ia at $1 < z < 2$, and to perform
the follow-up observations necessary for such a study, requires observing
deeper than can be feasibly done with the ground-based telescopes. However,
with the {\it Hubble Space Telescope (HST)} and the Advanced Camera for Surveys
(ACS), a higher-$z$ SN survey is practical. Through careful planning, the Great
Observatories Origins Deep Survey (GOODS) has been designed to accommodate a
deep survey for SNe with a specific emphasis on the discovery and follow-up of
$z \ga 1$ SNe~Ia.

We discovered 42 SNe over the 8-month duration of the survey. We also measured
redshifts, both spectroscopic and photometric, for all but one of the SN host
galaxies. For the first time, we have a significant sample of SNe~Ia spanning
a large range in redshift, from a complete survey with well understood
systematics and limitations. Certainly this has allowed for precise measurement
of the SN rates and the rate evolution with redshift~\citep[See][]{Dahlen2003},
but it also allows for a comparison of the observed SN~Ia rate history to the
star-formation rate history, and thus an analysis of SN~Ia assembly time, or
``delay time,'' relative to a single burst of star formation. By exploring the
range and distribution of the time from progenitor formation to SN~Ia explosion
that is required by the data, we can provide clues to the nature of the
mechanism (or mechanisms) which produce SNe~Ia.

We describe the Hubble Higher-$z$ Supernova Search (HHZSS) project in
\S~\ref{sec:GOODS}, along with image processing and reduction, transient
detection, and SN identification methods. In \S~\ref{sec:results} we show the
results of the survey, including discovery information on all SNe, and
multi-epoch, multi-band photometry of SNe over the search epochs of the
survey. In \S~\ref{sec:dt} we report on observational constraints on the
inherent SN frequency distribution, or the distribution ``delay times'' for
SN~Ia progenitors, and discuss the implied constraints on the model SN~Ia
progenitor systems. Elsewhere, we report on the rates of SNe~Ia and
core-collapse SNe, the comparison of these measured rates to those made by
other surveys, and to the predicted SN formation-rate history partly predicted
from the analysis in this paper~\citep{Dahlen2003}. In another paper we report
on the constraints of cosmological parameters and the nature of high-$z$
SNe~Ia~\citep{Riess2003b}.

\section{GOODS and the ``Piggyback'' Transient Survey}\label{sec:GOODS}

\begin{figure}
	\plotone{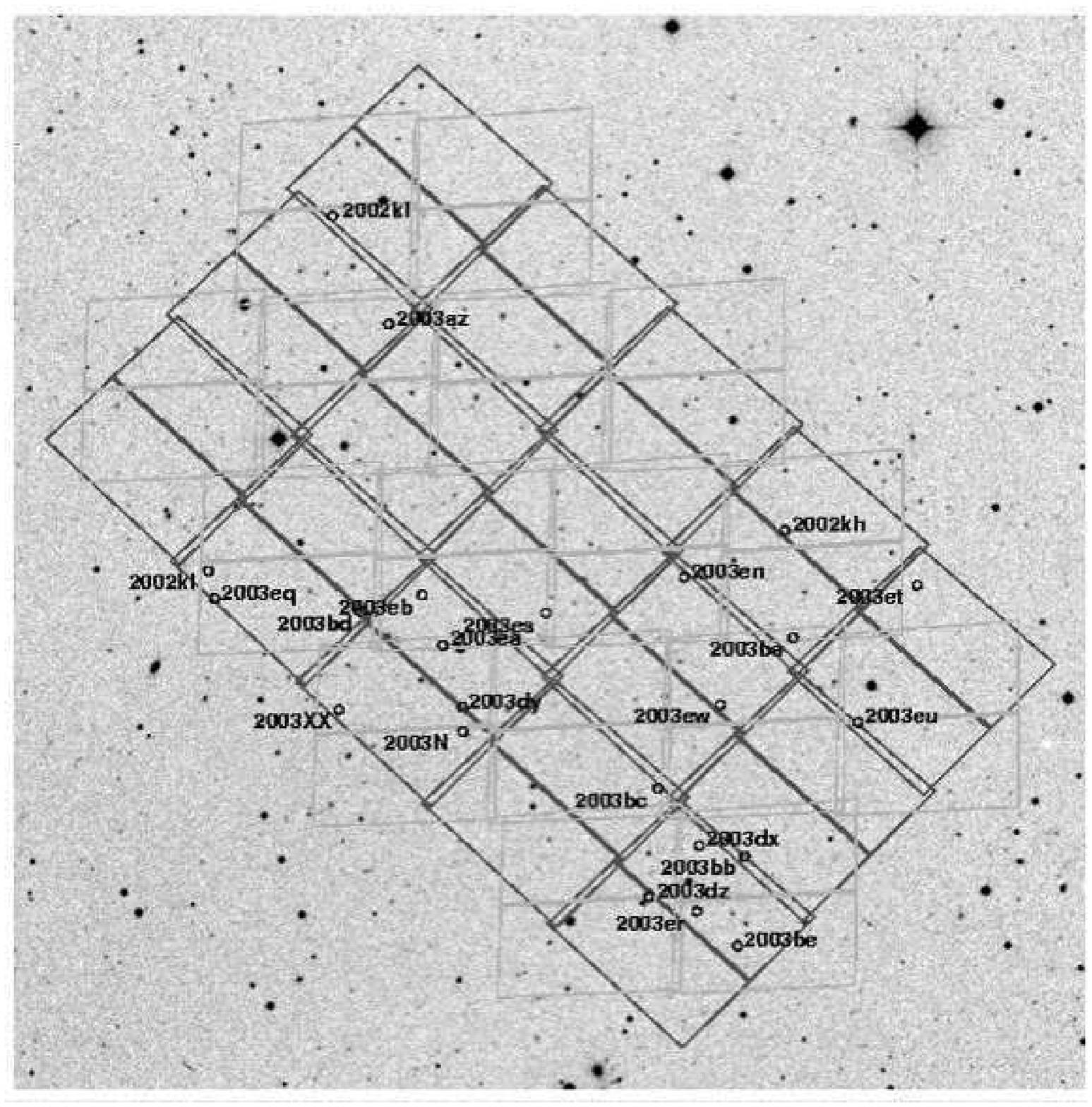}
	\caption{HDFN field observed by the GOODS project. North is up, and east is to the left. The tiles show the ACS
pointings for the first (dark) and second (grey) epochs. Epochs 3 and 5 are 
rotated by $90^\circ$ and $180^\circ$ (respectively) relative to epoch 1. 
Epoch 4 is rotated by $90^\circ$ relative to epoch 2. The SNe discovered in 
this field are marked and labeled.}
	\label{fig:fieldhdfn}
\end{figure}

\begin{figure}
	\plotone{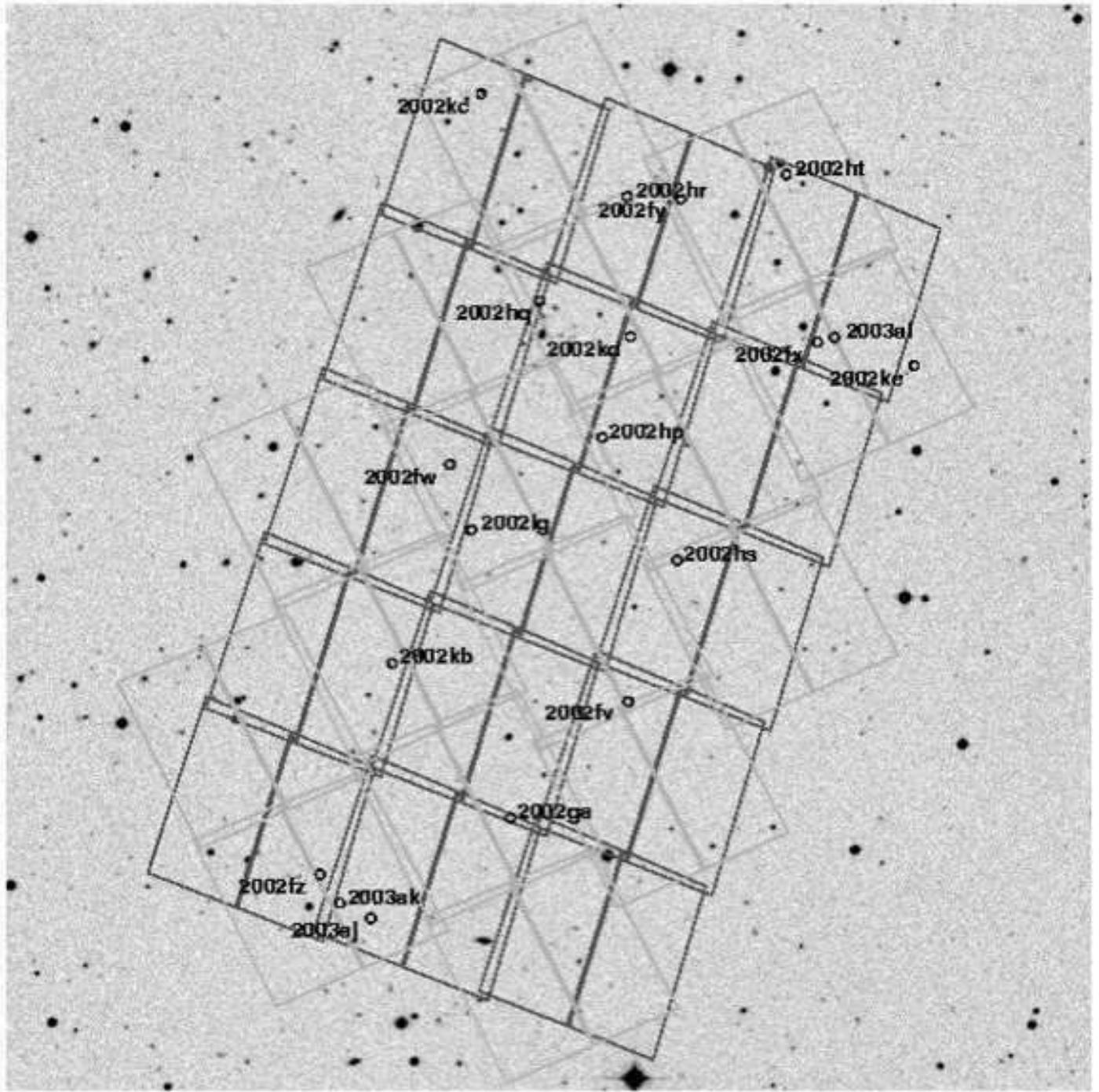}
	\caption{Same as in Figure~\ref{fig:fieldhdfn}, but for first (dark) and second (grey) epochs the CDFS field. Subsequent epochs are rotated by the same amounts as indicated in Figure~\ref{fig:fieldhdfn}.}
	\label{fig:fieldcdfs}
\end{figure}

GOODS was designed to combine extremely deep multi-wavelength observations to
trace the galaxy formation history and the nature and distribution of light
from star formation and active nuclei~\citep{Giavalisco:2003ig}. Using {\it
HST}/ACS, it has probed the rest-frame ultraviolet (UV) to optical portion of
high-redshift galaxies through observations in the $F435W$, $F606W$, $F775W$,
and $F850LP$ bandpasses, with a goal of achieving extended source sensitivities
only 0.5--0.8 mag shallower than the original Hubble Deep Field
observations~\citep{1996AJ....112.1335W}. Images were obtained in 15
overlapping ``tiled'' pointings, covering a total effective area of $\sim 150$
square arcmin per field. Two fields with high ecliptic latitude were observed,
the Chandra Deep Field South (CDFS) and the Hubble Deep Field North (HDFN), to
provide complementary data from missions in other wavelengths ({\it Chandra}
X-ray Observatory, {\it XMM-Newton}, {\it Spitzer} Space Telescope) and to
allow ground-based observations from both hemispheres (see
Figures~\ref{fig:fieldhdfn} and \ref{fig:fieldcdfs}).

The GOODS observations in the $F850LP$ band were scheduled over 5 epochs
separated by $\sim 45$ days to accommodate a ``piggybacking'' transient survey.
This baseline is ideal for selecting SNe~Ia near peak at $z \approx 1$, and SNe~Ia
on the rise at $z > 1.3$, as the risetime (from explosion to maximum brightness)
for SNe~Ia is $\sim 20$ days in the rest frame~\citep{Riess19991182675R}. The
baseline also insures that no SN in the desired redshift range will have sufficient time to rise
within our detection threshold, and then fall beyond detection before the field
is revisited, maximizing the overall yield.

Intentionally, the GOODS filter selections were nearly ideal for the detection,
identification, and analysis of high-redshift SNe~Ia. For a SN~Ia at $z \approx
1$, the $F850LP$ band covers nearly the same part of the SED as the rest-frame
$B$ band. The K-correction, or the correction of the observed flux to some
rest-frame bandpass (e.g., $F850LP$ to $B$), is thus relatively small.

Monte Carlo simulations of the survey, assuming detection limits based on the
$\sim 2100$~s exposure times per epoch (using the ACS Exposure Time Calculator)
and the desired baseline between epochs, implied that the distribution of
SNe~Ia would be centered at $z \approx 1$, with $\sim 1/3$ to $1/2$ of the
events occurring in the $1 < z < 2$ range. Scaling from other lower-$z$ SN
survey yields, it was expected that a total of 30--50 SNe of all types would be
discovered, and that $\sim 1/2$ of them would be SNe~Ia. These numbers implied
that we could expect to find $\sim 6$ to 8 SNe~Ia in the range of $1.2 \la z
\la 1.8$, which could be sufficient for an initial investigation of cosmology
in the deceleration epoch.

\subsection{Image Processing and Search Method}

The success of this survey has been due, in large part, to the rapid processing
and delivery of data, and the rapid post-processing by a reliable pipeline. The
exposures constituting a single tile in a single passband arrived from {\it
HST} within 6 to 18 hours after observation (with an average of $\sim 10$
hours), and fully processed (differenced with previous epochs) within a few
hours after arrival. In general, the complete multi-wavelength data for a
single tile were fully searched for candidate SNe within a day after
observation.

The individual exposures of a tile in a given epoch were reduced
(bias-subtracted and flat-field corrected) through the {\tt calacs} standard
ACS calibration pipeline. The well-dithered subexposures (or CR splits; see
below) were then corrected for geometric distortions and combined using the
{\tt multidrizzle} pipeline~\citep{koekemoe2003}. For the survey, we maintained
the physical pixel size of $0.05''$ pixel$^{-1}$ for the discovery of
transients.

A key feature of this pipeline is its identification and removal of cosmic rays
(CRs) and hot pixels. Each 2100~s exposure in $F850LP$ consisted of 4 individual
520~s CR splits, each dithered by small offsets. In each of the CR splits, the CR
contamination, at the time of the survey, was as high as $\sim 1\%$ of all
pixels, and hot pixels accounted for an additional $\sim
1\%$~\citep{riess2002}. With such a high incidence of CRs and hot pixels,
averaging (or taking the median) over the few CR splits would not adequately
remove these potential confusion sources. Instead, we used the {\tt minmed}
algorithm described in~\citet{acsdatahb}. Basically, of the pixels in each CR
split covering the same area of sky, the highest-value pixel was rejected. The
median of the remaining three pixels was then compared to the minimum-value
pixel. If the minimum pixel was within 6$\sigma$ of the median, then the median
value was kept, otherwise the minimum value was used. A second pass was
performed, repeating the {\tt minmed} rejection on pixels neighboring those
which had been previously replaced with minimum values (indicating CR or
hotpixel impact), but at a lower threshold to remove ``halos'' around bright
CRs. The result was that each pixel of the output combined image was either the
median or the minimum of the input values. Admittedly, the combined result was
less sensitive than can be obtained in a straight median, but the {\tt
multidrizzle} algorithm (with {\tt minmed}) did successfully reject $>99\%$ of
CRs and hot pixels after combination.

The search was conducted in 8 campaigns (4 campaigns for each of the HDFN and
the CDFS surveys) by differencing images from contiguous epochs. For a given
tile in a field, images covering the same area from the previous epoch were
aligned (registered) using the sources in the tiles. Catalogs of the pixel
centroids and instrumental magnitudes of sources on each image were made using
{\tt SExtractor}~\citep{Bertin1996393} and fed into a triangle matching routine
({\tt starmatch}, courtesy of B. Schmidt) which determined the linear
registration transformation from one epoch to the next. The typical precision
of the registrations was 0.2--0.3 pixels root-mean square (RMS), and the
point-spread function (PSF) in each epoch of observation remained nominally at
0.10--$0.13''$ FWHM. The combination of precise registration and nearly
constant PSF allowed for images to be subtracted directly, without the need for
image convolution.

\begin{figure}
	\epsscale{0.8}
	\plotone{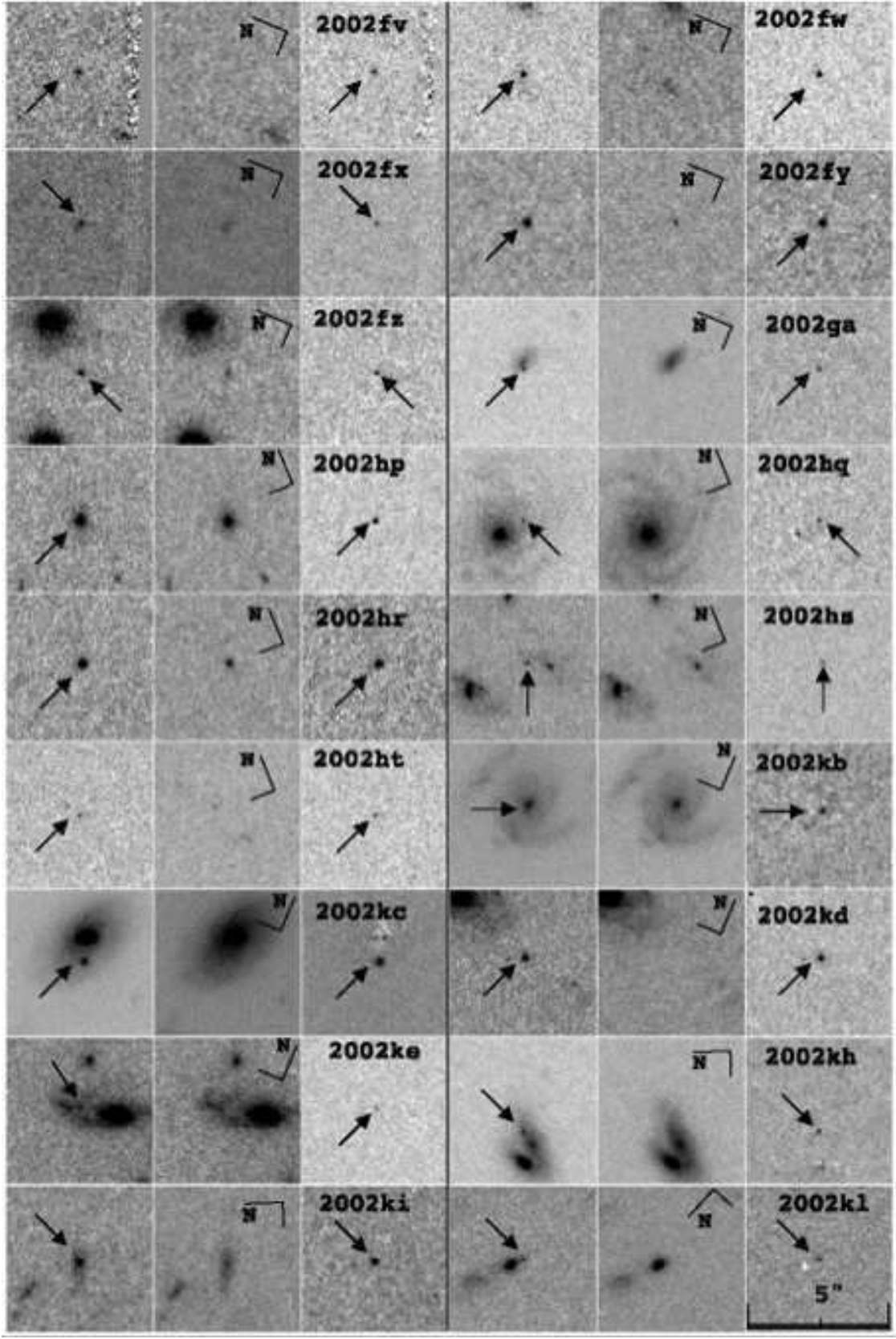}
        \caption{Discovery images for SN 2002fv through SN 2002kl. Each SN has
three panels: the discovery image (left), a template constructed from images
without the SN (middle), and the subtraction of the two (right). The SN is
labeled in the subtraction image. Arrows indicate the position of each SN in 
the discovery and subtraction images. North and east are marked. The image 
scale is shown in the lower right-most image.}

	\label{fig:imsub1}
\end{figure}

\begin{figure}
	\epsscale{0.8}
	\plotone{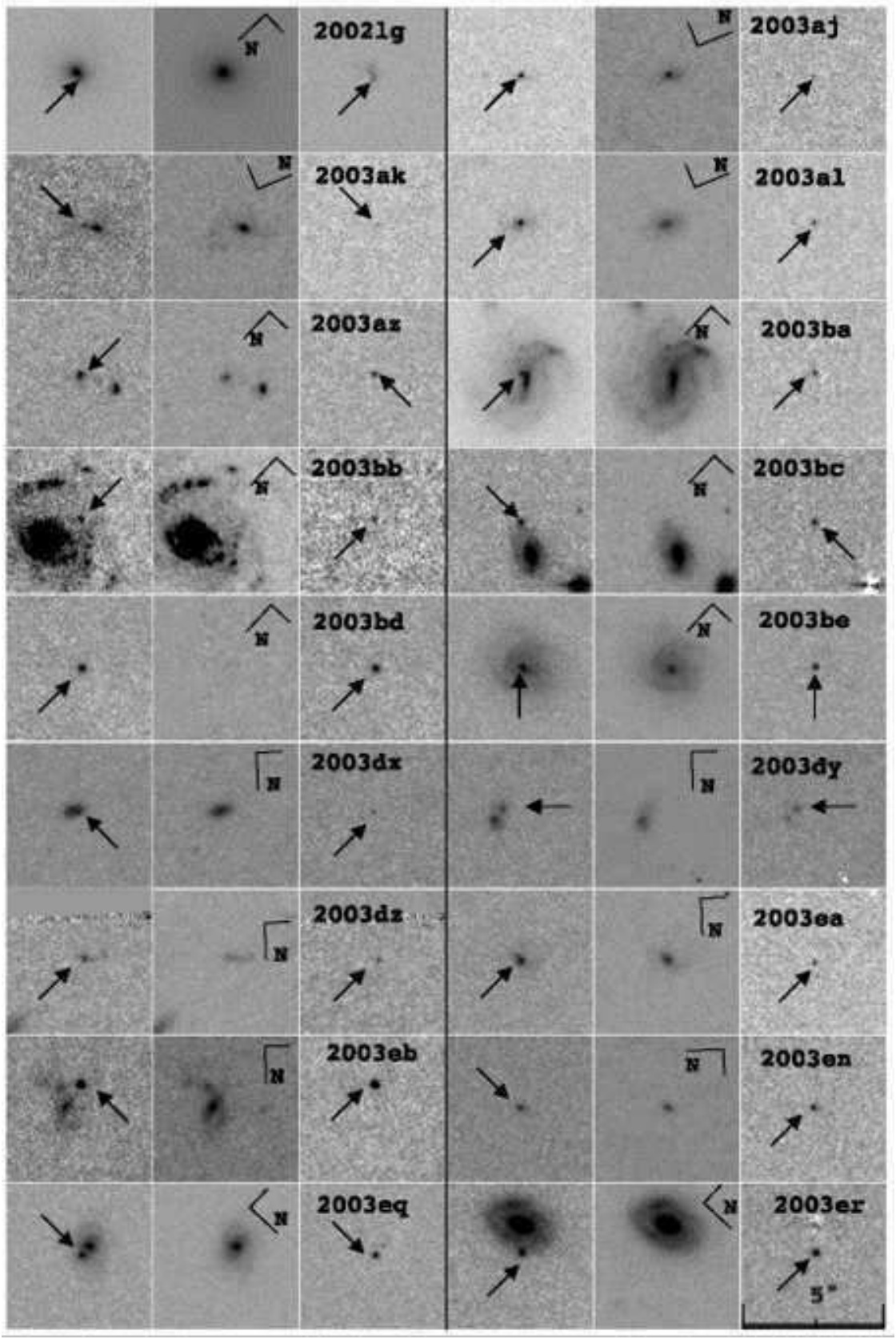}
	\caption{Same as in Figure~\ref{fig:imsub1}, but for SN 2002lg  
through SN 2003er.}
	\label{fig:imsub2}
\end{figure}

\begin{figure}
	\epsscale{0.8}
	\plotone{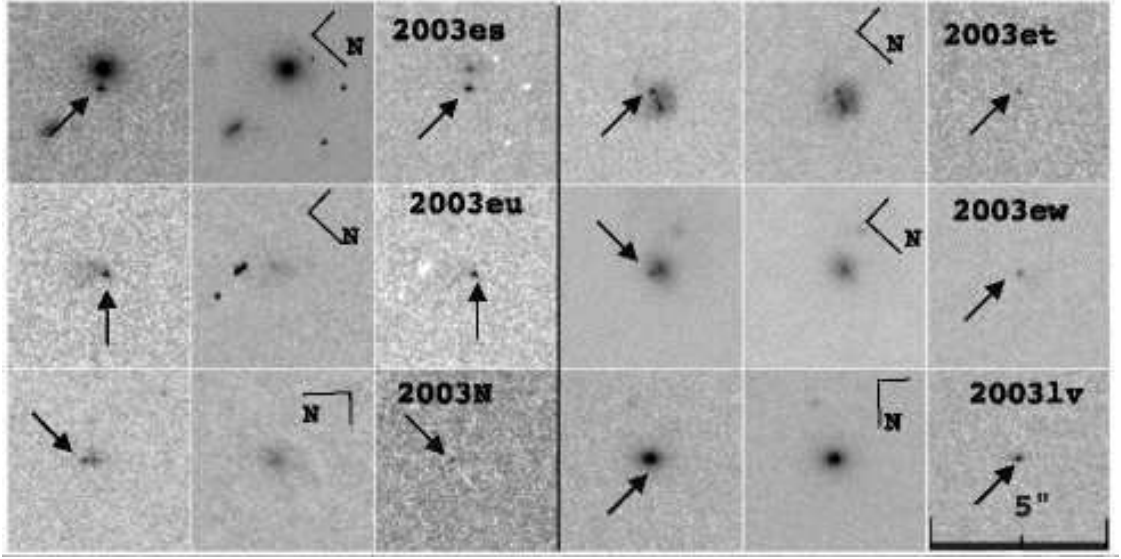}
	\caption{Same as Figure~\ref{fig:imsub1}, but for SN 2003es through 
SN 2003lv.}
	\label{fig:imsub3}
\end{figure}

Several examples of the image subtraction quality are shown in
Figures~\ref{fig:imsub1},~\ref{fig:imsub2},~and \ref{fig:imsub3}.  In ideal
situations, only transient sources remain in the residual image on a nearly
zero-level background. However, in practice there were many situations which
produced non-transient residuals. Although extensive care was taken to remove
many CRs and hot pixels in the image processing, these artifacts did
occasionally slip past the rejection algorithms, specifically when multiple
effects were coincident on the same area of sky. For example, for a given pixel
in each of the four CR splits covering the same area of sky, the probability
that the pixels were impacted by a CR in 3 of the 4 exposures is
approximately 1 in $10^6$. Roughly 20 pixels in the combined 20 million
pixel array would show CR residuals after passing through the {\tt
multidrizzle} algorithm.  In addition, ``breathing" in the optical path, focus
drift, and the slight change in the pixel scale across the image plane have all
led to small yet detectable variations in the PSF. Sometimes bright compact
objects were over-subtracted in the wings of their radial profiles and
under-subtracted in the inner 1--2 pixels. Other instrumental sources of
confusion include diffraction spikes, correlated noise from multiple image
resampling, and slight registration errors due to the lack of sources over a
large registration area.

The non-trivial abundance of false positives required rigorous residual
inspection methods. We therefore searched the subtracted images redundantly to
minimize false detection biases and to maximize recovery of elusive, faint
transients.  An automated routine was performed to identify PSF-like residuals
which were well separated ($\ge 2$ pixels) from known saturated pixels, and
above $\sim 4-5\sigma$ of the sky background. The inherent nature of this
routine prohibits the detection of nonstellar residuals, faint residuals,
residuals near bright stars or nuclei (which may be saturated), or residuals in
areas where the RMS of the background could not be easily determined by the
automated routine. Therefore, so that no potential SNe were lost, several human
searchers visually inspected each subtracted image. At least two pairs of
searchers {\it independently} scoured a few residual tiles. Visual searching of
only a few tiles insured that it was done thoroughly, and helped to alleviate
monotony and fatigue.

Candidate SNe found by the software and the searchers were then scrutinized
based on the following set of criteria to select SNe and further reject
instrumental (and astronomical) false positives:

 {\it (1) Misregistration:} Areas with $\lesssim 10$ detectable sources per
arcmin$^{2}$ are typically poorly registered ($\ga 0.5$ pixel RMS). Sources in
these areas of the subtracted images are under-subtracted on one side, and
over-subtracted on the other. If the total flux in an aperture encompassing the
source was not significantly greater than a few times the background RMS, it
was assumed that the residual was an artifact of misregistration.

 {\it (2) Cosmic ray residuals:} The number of pixels in $\sim 2100$~s combined
images that still contain CRs due to impacts on the same regions of sky on one
or more individual $\sim 520$~s exposures is roughly $4500^{2} \times
(0.01)^{N}$, where $N$ is the number of impacted exposures (out of 4). This
number can grow slightly when considering hot pixels and bright pixels with
CRs. To further reject these artifacts, we required that candidates have no
more than one constituent exposure affected by CRs or hot pixels.

{\it (3) Stellar profile:} Residuals in the subtracted images were required to show
a radial profile consistent with the PSF ($\sim 2$ pixels FWHM). Narrower
profiles were considered to be stacked noise (if not residual CRs) and wider
profiles were typically poor subtractions from misregistrations, breathing, or
focus drift.

{\it (4) Multiple epochs of detection:} It was required that each candidate be
detected (to within 5$\sigma$) on each of the CR split exposures that were not
impacted by CRs or hot pixels at the relevant location. Additional weight was
given to candidates that were clearly detected in the $F775W$ band, or
additionally in the $F606W$ band. However, this was not a strict criterion as
it was expected that SNe~Ia at higher redshifts would become less detectable in
the bluer wavelengths (see Section~\ref{sec:snid}).  

{\it (5) Variable galactic nuclei:} Sources that were $\la 1$ pixel from their host nuclei
were considered potential active galactic nuclei (AGNs) and typically not included in the target of opportunity follow-up program (see \S\ref{sec:follow}). However, these residuals were followed over subsequent search epochs, and in all but one case, sufficient photometric evidence (see \S\ref{sec:snid}) was found to classify them as SNe. Bright residuals that were coincident with
the nuclei of galaxies were also compared with known X-ray sources from the Chandra
Deep Field South and Chandra Deep Field North 1 Megasecond
catalogs~\citep{2001AJ....122.2810B,2002ApJS..139..369G}.  Indeed, the only variable source unidentified by spectroscopic or photometric means was identified as known X-ray source, and therefore rejected as the only confirmed optically variable AGN in the survey.

{\it (6) Solar-system objects and slow-moving stars:} We required that our
candidates show no proper motion. Assuming we were sensitive to 1/2-pixel
shifts, the proper motion of any candidate could not be more than $0.025''$
over the $\sim 2100$~s combined exposure, or $\omega < 0.043''$ hr$^{-1}$ (0.1
deg yr$^{-1}$).  Hypothetically, if a source was bound to the Sun (with
tangential velocity $\sim30$ km hr$^{-1}$), then its distance would have to be
$D>$ [(30 km s$^{-1}$)/(0.043'' hr$^{-1}$)], or greater than $3,400$ AU. In
addition, if the object was illuminated by reflected sunlight, then its
apparent magnitude ($m$) would be related to its angular diameter ($\theta$) by
\begin{equation} m=m_{\odot}+5\log(\theta/2D), \end{equation} where $m_{\odot}$
is the apparent magnitude of the Sun. Since $\theta$ must be consistent with the
PSF ($\sim 0.1''$), the source would have to be $\sim4$ times larger than Jupiter
(at the distance assumed from the limits on proper motion), and the apparent
magnitude of the source would have to be $m\approx55$ mag!  Alternatively, using the
limiting magnitude for the survey, $m_{\lim} \approx 26$ (see
\S~\ref{sec:controltime}), the source would have to have an angular size of
$\theta > 18^\circ$ in order to have been lit by the Sun at its assumed
distance.
	
A similar argument can be made for slow-moving stars. Since $\omega=0.043''$
hr$^{-1}$ is the fastest a source could move without being detected, in the
$\sim45$ days since the field was last observed the source could have moved $<
1000$ pixels. Our survey was clearly sensitive to negative residuals as well as
positive ones (a fact indicated by the frequent discovery of SNe declining in
brightness since the previous epoch). We saw no negative candidates which were
detected within 1000 pixels of a positive source on the same epoch of
observation.
	 
Most of these SNe have been observed on more than one epoch, and all but two
were detected within $3.5''$ of a galaxy (presumably the host). It would be
highly unlikely for any of these to be objects moving within the Solar
System or the Galaxy.

\subsection{Identification of SNe and Redshift Determination}\label{sec:snid}

SNe are generally classified by the presence or absence of particular features
in their optical spectra (see
\citeauthor{Filippenko1997309}~\citeyear{Filippenko1997309} for a
review). Historically, the primary division in type has been by the absence
(SNe~I) or presence (SNe~II) of hydrogen in their spectra, but the
classification currently extends to at least 7 distinct subtypes (SN IIL, IIP,
IIn, IIb, Ia, Ib, and Ic). It is now generally accepted that the explosion
mechanism is a more physical basis by which to separate SNe.  SNe~Ia probably
arise from the thermonuclear explosion of carbon-oxygen white dwarf stars,
while all other types of SNe are produced by the core collapse of massive stars
($\ga 10M_{\odot}$).

There can be considerable challenges in the ground-based spectroscopic
identification of high-redshift SNe. As the principal goal of this survey has
been to acquire many SNe~Ia at $z > 1$, a fundamental prerequisite was that we
could make confident identifications of at least this SN type. Much to our
benefit, {\it HST} with the ACS $G800L$ grism provides superb spectra with
significantly higher signal-to-noise ratio (S/N) than can currently be achieved
from the ground. Its limitation is the low spectral resolution ($R \equiv
\lambda/\Delta\lambda \approx 200$ per pixel, in first order) and the overlap
of multiple spectral orders from other nearby sources. Spectral resolution of
$\sim 1500$ km s$^{-1}$ is not problematic for SNe with ejecta velocities of
$\ga 10,000$ km s$^{-1}$. However, because of the spectral-order confusion and
the lack of a slit mask, the $G800L$ grism could only be used for SNe with
substantial angular separation from their hosts and from other nearby sources.

It was expected that SN candidates would generally be either too faint to be
spectroscopically observed from the ground, or too close to their host galaxies
or other nearby sources to be identified with the ACS grism. We therefore had
to rely on some secondary method by which to identify SNe, specifically to
select likely SNe~Ia from the sample. The inherent differences in the ejecta
compositions of SNe~Ia and SNe~II leads to an observable difference in their
intrinsic early-time UV flux. As optical observations shift to the rest-frame
UV for $z \gtrsim 1$ SNe, the ``UV deficit'' in SNe~Ia can be a useful tool for
discriminating SNe~Ia from SNe~II, the most common types of core-collapse (CC)
SNe. Using a method pioneered by~\citet{Panagia03} and fully developed
in~\citet{Riess2003b}, we use the $F850LP$ apparent magnitude, the
$F775W-F850LP$ and $F606W-F850LP$ colors, the measured redshift or photometric
redshift estimates (see below), and age constraints provided by the baseline
between search epochs to grossly identify SNe as either SNe~Ia or SNe~CC. This
method is only useful for $z \ga 1$ SNe near maximum light, and is not
foolproof in its identification. There are SNe~CC (e.g., luminous SNe~Ib and
Ic) which can occupy nearly the same magnitude-color space as SNe~Ia. However,
these bright SN~Ib/c make up only $\sim 20$\% of all SNe~Ib/c, which as a group
are only $\sim 1/3$ as plentiful as other SNe~CC~\citep{1999AA...351..459C}.

From the ground, we have obtained spectroscopic identification of 6 SNe~Ia and
1 SN~CC in the redshift range 0.2--1.1 using Keck + LRIS (see
Table~\ref{tab:sntable}). With {\it HST}/ACS and the $G800L$ grism, we have
obtained excellent spectra of 6 SNe~Ia at $z = 0.8$--1.4, the most distant
sample of spectroscopically confirmed SNe; see~\citet{Riess:2003gz}.  These
spectra cover only the 2500--5000~\AA\ range in the rest frame, but they are of
excellent S/N, unattainable for such high-$z$ SNe from the ground. These
identifications also serve as an excellent proof of concept in the
color-magnitude selection.

\begin{figure}
	\centering
 	\includegraphics[angle=-90,width=\textwidth]{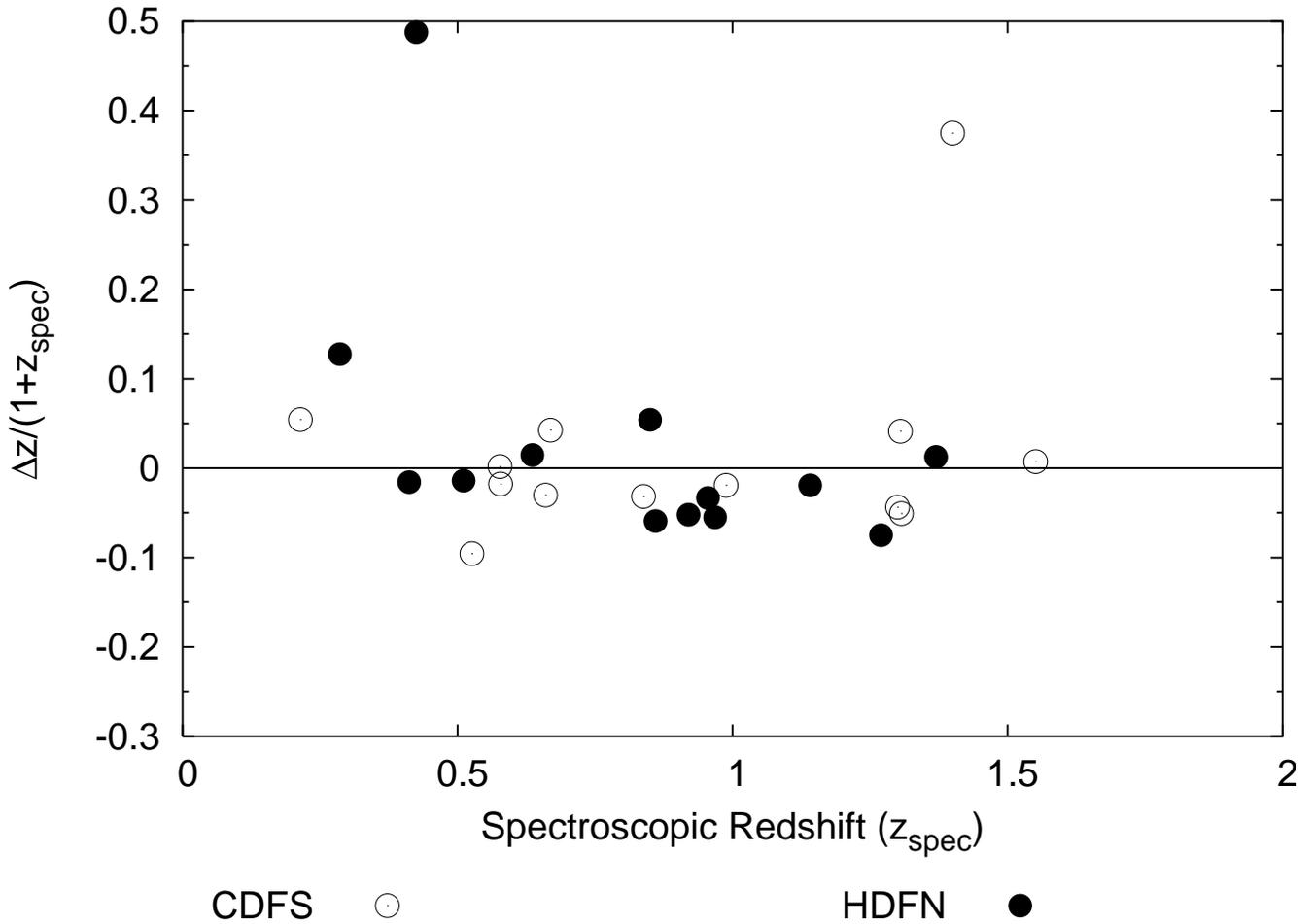}
        \caption{The accuracy of the photometric redshifts as a function of actual
spectroscopic redshift for the 26 SN host galaxies. Photometric redshifts were
precise to $\sim0.05$ RMS (rejecting two $>7\sigma$ outliers).}
	\label{fig:photz}
\end{figure}

Using Keck, the VLT, and the ACS grism, we have obtained spectroscopic
redshifts for 29 of the 42 SNe in our sample. To our benefit, part of the GOODS
endeavor involved obtaining extensive multi-wavelength photometry spanning the
$U$ to the near-IR passbands to estimate the photometric redshifts
(``phot-$z$") of galaxies in the HDFN and CDFS
fields~\citep{Mobasher:2003bm}. The precision of the phot-$z$ from GOODS with
respect to known spectroscopic redshifts has been within $\sim 0.1$ RMS, with
the occasional instance ($\sim 10\%$ of a tested sample) where the phot-$z$
method misestimates the actual redshift by more than 20\%. In order to improve
on the accuracy of the phot-$z$ measurements for the host galaxies, we remeasured the multi-wavelength photometry by visually determining the centroid of the host galaxies, and manually determining an annulus in which the sky background is determined. This allowed better photometric precision than was generally achieved in the {\it SExtractor}-based automated cataloging. Comparing the sample of 26 SN host galaxy spectroscopic redshifts
to the phot-$z$ estimates from the improved photometry\footnote[16]{Only 26 SN
host galaxies have both measured spectroscopic and photometric redshifts.}
resulted in a precision of 0.05 RMS (after rejecting two $>7\sigma$ outliers),
and only $\sim 5\%$ of the sample was misestimated by more than 10\% (see
Figure~\ref{fig:photz}). The redshifts of the remainder of the SN hosts
(without spectroscopic redshifts) were determined in this way, with the
exception of SN~2002fv, whose host was not identified due to the magnitude
limits of the survey.

We fit template light curves to grossly identify SNe which were not
spectroscopically identified, and were not at $z \ga 1$ nor constrained near
maximum light. Using the light curves of SNe~1994D, 1999em, 1998S, and 1994I as
models for SNe~Ia, IIP, IIL, and Ib/c (respectively), we transformed these
model SNe to the redshifts of the observed SNe, correcting for the effects of
time dilation, and applying K-corrections to the rest-frame bandpasses to
produce light curves as they would have been seen through the $F850LP$,
$F775W$, and $F606W$ bandpasses at the desired redshifts. The K-corrections
were determined from model spectra~\citep{2002PASP..114..803N} for SNe~Ia, and
from color-age light-curve interpolations for SNe~CC. We have also made use of
the web tool provided by~\citet{2002PASP..114..833P} to check the derived
colors for the SNe~CC. We visually determined the best-fit model light curve to
the observed light curves, allowing shifting along the time axis, magnitude
offsets, and extinction/reddening (assuming the Galactic extinction law) along
the magnitude axis. Best fits required consistency in the light-curve shape and
peak color (to within magnitude limits) and in peak luminosity in that the
derived absolute magnitude in the rest-frame $B$ band had to be consistent with
the observed distribution of absolute $B$-band magnitudes shown
in~\citet{2002AJ....123..745R}.

Each discovered SN was given an identity rank (gold, silver, or bronze)
reflecting our confidence in the identification.  A gold rank indicated the
highest confidence that the SN was the stated type, and it was not likely that
the SN could have been some other SN type. A silver rank indicated the identity
was quite confident, but the SN lacked sufficient corroborating evidence to be
considered gold. A bronze rank indicated that there was evidence the SN type
was correct, but there was a significant possibility that the SN type was
incorrect.

We were clearly confident of the SN type in cases where a high S/N ($\ga 20$)
spectrum conclusively revealed its type; these SNe were gold, by
definition. However, the majority of SNe were without sufficient spectra to
unambiguously determine a type.  We then used additional information on the SN
redshift, photometric data, and host-galaxy morphology, seeking a consistent
picture for a specific SN type.

We first considered the possibility that a candidate was a SN~Ia. We required
that the light-curve shape was at least consistent with a SN~Ia at its
redshift, and that the observed colors and derived absolute magnitude could be
made consistent with the template light-curve colors with $< 1$ mag of
extinction (assuming the Galactic extinction law). If the SN was at $z \ga 1$, and its peak colors were $F775W-F850LP \ga 0.5$ mag and $F606W-F850LP \ga 1$ mag, we considered it highly likely to be a SN~Ia.

The study of~\citet{2000AJ....120.1479H} has shown that at low redshifts,
early-type galaxies (ellipticals) only produce SNe~Ia, and have not as yet been
shown to produce SNe~CC. Hence, we regard SNe found in red elliptical hosts to
have been most likely SNe~Ia and unlikely SNe~CC.

Based on the above information, any SN in our survey at $z > 1$, in a red
elliptical host, and having light curves and peak colors consistent with a SN~Ia,
were most confidently considered SNe~Ia and ranked ``gold SNe~Ia." SNe having
photometric data consistent with SNe~Ia, and either at $z > 1$ (identifiable
by their peak color) or in early-type host galaxies, were considered SNe~Ia
with a high confidence, and therefore ranked ``silver SNe~Ia." SNe having light
curves consistent with SNe~Ia, but without any other information to confirm their
type, were ranked as ``bronze SNe~Ia."

\begin{figure}
	\plotone{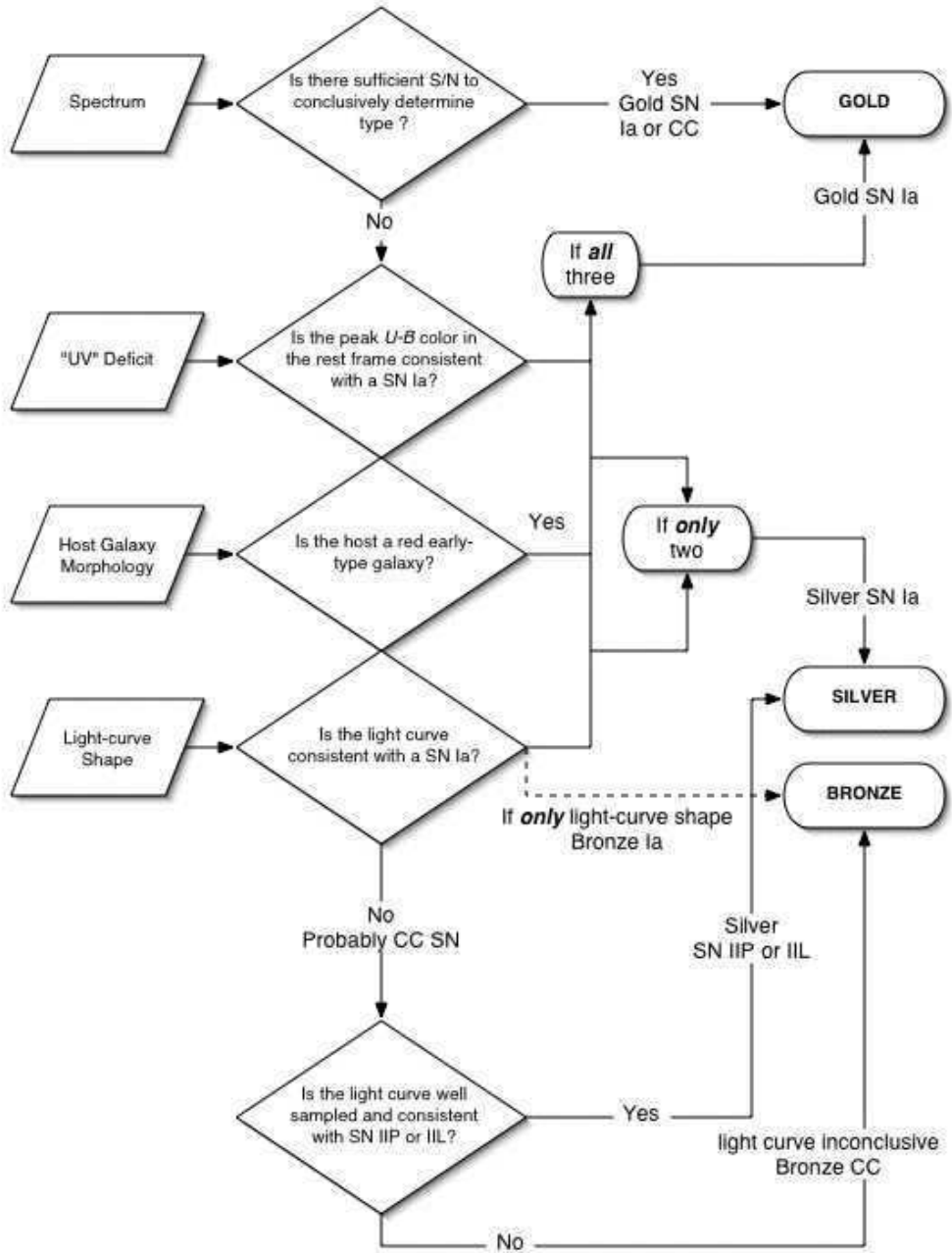}
	\caption{Flow chart showing how SN types and confidence ranks were 
determined from the data.}
	\label{fig:flow}
\end{figure}

If the light curves for a SN seemed inconsistent with a SN~Ia, we compared them
to the model light curves for SNe~CC. If the SN showed a slow rate of decline
from peak (consistent with SNe~IIP and some SNe~IIn), then it was considered a
SN~CC with high confidence, or a ``silver SN~CC." All other SNe, inconsistent
with SNe~Ia, SNe~IIP, or slowly declining SNe~IIn, were placed into the
``bronze SN~CC" category. For clarity, we include a flow chart showing the
conditions used to determine the identification confidence rank
(Figure~\ref{fig:flow}).

\subsection{Follow-up {\it HST} Observations}\label{sec:follow}

An intensive target of opportunity (ToO) follow-up program with {\it HST} (GO
9352; Riess, PI) was conducted for candidate SNe~Ia in the range of $z \ga
1$. The decision to trigger the ToO was based on the prior certainty of SN~Ia
type and redshift range. These observations are intended to support
multi-wavelength light-curve shape fitting~\citep{RPK199647388} with multiple
observations in passbands as close to the rest-frame $U$, $B$, and $V$ bands as
possible. The ToO program consisted of supplementary observations with ACS (in
$F775W$ and $F850LP$ bands), NICMOS (in the $F110W$ and $F160W$ bands), and ACS
$G800L$ grism spectra when feasible. These observations were rapidly initiated
(within $\sim1$ week of SN detection) so that identification and color
measurements could be made as near to maximum light as possible, and so that
the light-curve sampling could be optimized. Using an updated version of the
multicolor light-curve shape algorithm (Jha, Riess, \& Kirshner 2004, in
preparation), we estimate key parameters of the rest-frame optical light curves,
particularly the $B$-band magnitude at maximum, the rest-frame $U-B$ and $B-V$
colors at maximum, and the rate of decline from maximum light in the $B$
band. Further details on the ToO program, including the photometric and
spectroscopic data, can be found in~\citet{Riess2003b}.

\section{Results of the Survey}\label{sec:results}

Over the course of 8 search campaigns from 2002 August to 2003 May, we
successfully discovered 42 SNe of both physical types over a wide range of
redshifts. The SNe are shown in their discovery-epoch images in
Figures~\ref{fig:imsub1},~\ref{fig:imsub2}, and \ref{fig:imsub3}. They are
listed in Table~\ref{tab:sntable} with their U.~T. date of discovery, coordinates, physical SN type,
type confidence, redshift, source of measured redshift, and offset from host
galaxies, if detected.

The optical {\it HST}/ACS photometry for most of the SNe is given in
Table~\ref{tab:snphot}. This photometry consists of discovery epoch apparent
magnitudes, and data on the SN (when detected) from subsequent search
epochs. The optical and infrared photometry for the 16 SNe~Ia which were used
in the cosmological analysis (specfically SNe 2002fw, 2002fx, 2002hp, 2002hr,
2002kc, 2002kd, 2002ki, 2003ak, 2003az, 2003bd, 2003be, 2003dy, 2003eb, 2003eq,
2003es, and 2003lv) are shown in~\citet{Riess2003b}. The data listed in Tables~\ref{tab:sntable} and~\ref{tab:snphot}, and in~\citet{Riess2003b}, supersede preliminary data announced in IAU Circulars  7981, 8012, 8038, 8052, 8069, 8081, 8083, 8125, 8140, and 8141.

For each SN, images from all survey epochs in which the SN was not detected (to
within a 10$\sigma$ limit) were combined to create a template image. Images from
each epoch in which the SN was detected had the template image subtracted from
it to remove the host galaxy and other background light. The apparent magnitude
in each passband was measured through a narrow aperture ($0.15''$ radius)
centered on the SN. The residual sky brightness (and noise) were determined in
larger aperture annuli (0.6--1$''$). Aperture corrections determined
by~\citet{2002hstc.conf...61G} were applied to correct from the encircled flux
in the narrow aperture to what would be expected in a nearly infinite
aperture. We then measured the apparent magnitudes relative to the 1 count per
second zero points determined by~\citet{2003SPIE.4854..496S}. Photometric
errors were approximated using the ACS Exposure Time Calculator.

\section{Delay Time Functions and Models for SNe~Ia Progenitor Systems}\label{sec:dt}

The current consensus on SNe~Ia is that they are thermonuclear explosions of
white dwarf (WD) stars as they accrete matter to reach the Chandrasekhar
mass~\citep{Livio99}. The two most likely scenarios are single degenerate (SD)
systems (a single WD accreting material from a normal companion star), and
double degenerate (DD) systems (the merger of two WDs). It is not yet fully
understood which scenario represents the preferred mechanism or channel for the
production of these events, or if more than one channel is used by progenitors
to make SNe~Ia.  To that end, there is some uncertainty concerning the
characteristic time scale from the formation of these progenitors to the
occurrence of the events, and concerning the distribution of these delay
times. Nevertheless, there is some consensus that the delay time in the SD
scenario is chiefly governed by the main-sequence lifetime of the companion
star which is on the order of 10$^{9}$ yr, and in DD by the time necessary to
gravitationally radiate away the angular momentum~\citep{1984ApJS...54..335I,
1994MNRAS.268..871T} which is on the order of 10$^{8}$ yr. Chemical evolution
in the solar neighborhood~\citep{1996ApJ...462..266Y}, and additional SD/DD
modeling ~\citep{1998ApJ...497L..57R,1999ApJ...522..487H}, suggest 0.5--3 Gyr
mean delay times should be plausible for SD, and a mean of $\sim 0.3$ Gyr in DD.

Even within the SD scenario, there is quite a diversity of specific models. In
addition to the substantial mass accretion [$\dot{M}_{acc} \approx (5 -
10)\times10^{-8}$ M$_{\odot}$ yr$^{-1}$], there can be significant winds
($\dot{M}_{wind} \approx -0.5 \dot{M}_{acc}$), and possibly companion-mass
stripping ($\dot{M}_{strip} \approx -0.1 \dot{M}_{acc}$) to accommodate a
larger range in companion star masses~\citep{1999ApJ...522..487H}. Indeed,
there are a variety of SD models which can reproduce a satisfactory set of
SN~Ia characteristics, but none as yet which have thoroughly accounted for the
SN~Ia diversity (see~\citeauthor{Livio99}~\citeyear{Livio99} for a review). It
is possible, in light of this diversity, that there are several channels by
which SNe~Ia are produced. For example, it is possible to imagine a scenario in
which SD channels account for the majority of SN~Ia events, perhaps $\ga80\%$,
and the other $\sim 20\%$ would come from DD systems. This would be consistent
with the observed luminosity diversity at $z \approx 0$, and, assuming some
simple evolutionary arguments, could account for the apparent lack of diversity
at higher redshifts~\citep{Livio99, Li2001ApJ546734L}. Ultimately, it is this
uncertainty in the progenitor systems that inevitably makes it difficult to
quantify the intrinsic distribution of delay times, which would allow a
comparison of the observed SN~Ia rate to the star formation rate.

We therefore attempt to constrain the apparent distribution of delay times
through the observed SN rates and measurements of the star formation
history. The frequency distribution, or number distribution [$N_{Ia}(z)$], of
SNe in our survey can be given by

\begin{equation}
	N_{Ia}(z)=\text{SNR}_{Ia}(z)\times t_{c}(z)\times(1+z)^{-1}\times\frac{\Theta}{4\pi}\times\Delta V(z),
	\label{eqn:numdist}
\end{equation}

\noindent where SNR$_{Ia}(z)$ is the intrinsic SN volume rate (number per unit
time per unit comoving volume). The survey's efficiency with redshift is
represented as a ``control time,'' $t_c(z)$, or the amount of time in which a
SN~Ia at a given redshift {\it could} have been observed by our survey (see
\S~\ref{sec:controltime}). $\Theta$ is the solid angle of the survey area
($\sim300$ sq. arcmin, or $2.54\times10^{-5}$ steradians). $\Delta V$ is the
volume comoving element contained in a shell about $z$ and is defined by
$\Delta V(z)\equiv V(z+\Delta z) - V(z)$, with

\begin{equation}
	\begin{split}	
	V(z)=&4\pi H_0^{-3}(2\Omega_k)^{-1}\\
	&\times\biggl\{H_0\biggl(\frac{D_L(z)}{1+z}\biggr)\biggl[1+\Omega_kH_0^2\biggl(\frac{D_L(z)}{1+z}\biggr)^2\biggr]^{1/2}\\&-\vert\Omega_k\vert^{-1/2}sinn^{-1}(H_o\biggl(\frac{D_L(z)}{1+z}\biggr)\vert\Omega_k\vert^{1/2})\biggr\},	\end{split}
	\label{eqn:vol}
\end{equation}

\noindent where $H_0$ is the Hubble constant at the present epoch $t_0$, and
$D_L$ is the luminosity distance. Here ``$sinn$'' and $\Omega_k$ are terms
which describe the curvature of space, where $sinn=sinh$ when $\Omega_k > 0$
(open Universe), and $sinn=sin$ when $\Omega_k<0$ (closed Universe)\footnote[17]{In
the case of $\Omega_k=0$, eq.~\ref{eqn:vol} becomes
$V(z)= (4\pi/3)D_L^3 (1+z)^{-3}$.}. $\Omega_k$ is defined by
$1-\Omega_k=\Omega_M+\Omega_{\Lambda}$. 

We assume that the intrinsic SN~Ia rate would be a reflection of the star
formation rate, SFR$(z)$, distorted and shifted to lower redshifts by the
convolved delay time distribution function, $\Phi(t_d)$:

\begin{equation}
	\text{SNR}_{Ia}(t)=\nu\int_{t_F}^{t}\text{SFR}(t')\times\Phi(t-t')~dt',
	\label{eqn:snriadef}
\end{equation}	

\noindent where
$\int\text{SNR}_{Ia}(z)~dz\equiv\int\text{SNR}_{Ia}(t)~dt$. Here, $t$ is the
age of the Universe at redshift $z$, $t_F$ is the time when the first stars
were formed, and for computational convenience, we set $z_F = 10$. We define
$\nu$ as the number of SNe~Ia per formed solar mass. Therefore, $\Phi(t_d)$ is
the frequency distribution of SNe~Ia (yr$^{-1}$), and represents the relative
number that explode at a time $t_d$ since a single burst of star formation.

As the HHZSS$-$GOODS data span a vast range in redshift extending to
$z \approx 1.6$, they are well suited to probing SNR$_{Ia}(z)$, and to
determining $\Phi(t_d)$. In this analysis, we attempt to determine
constraints on $\Phi(t_d)$ by testing a few model distributions in their
ability to recover the observed redshift distribution of SNe~Ia from this survey. Overall normalization factors, such as the number of SNe per unit formed stellar mass, are largely ignored in this analysis. The actual rates of SNe~Ia (including normalization) are calculated and analyzed in~\citet{Dahlen2003}. We use
the gold, silver, and bronze SNe~Ia together (a total of 25 SNe~Ia)
throughout this analysis, and we assume $\Omega_M=0.30$,
$\Omega_{\Lambda}=0.70$, and $H_0 = 70$ km s$^{-1}$ Mpc$^{-1}$.

\subsection{The Star Formation Rate Model}

Various observations of galaxies in the rest-frame passbands have given
information on SFR$(z)$, now extending to
$z\ga5$~\citep{Giavalisco:2003bi}. This current model broadly supports the
findings of~\citet{1998ApJ...498..106M} in suggesting that SFR$(z)$ is peaked
at $1 < z < 2$, but it is substantially flatter in its decline at $z > 2$. There is,
however, some uncertainty in the amount of correction for extinction in the
galaxies themselves
(see~\citeauthor{Giavalisco:2003bi}~\citeyear{Giavalisco:2003bi} for a
discussion). Indeed, without the extinction correction, the deduced SFR$(z)$
would be similar to the~\citet{1998ApJ...498..106M} function, but extending to
higher redshifts.

\begin{figure}
	\plotone{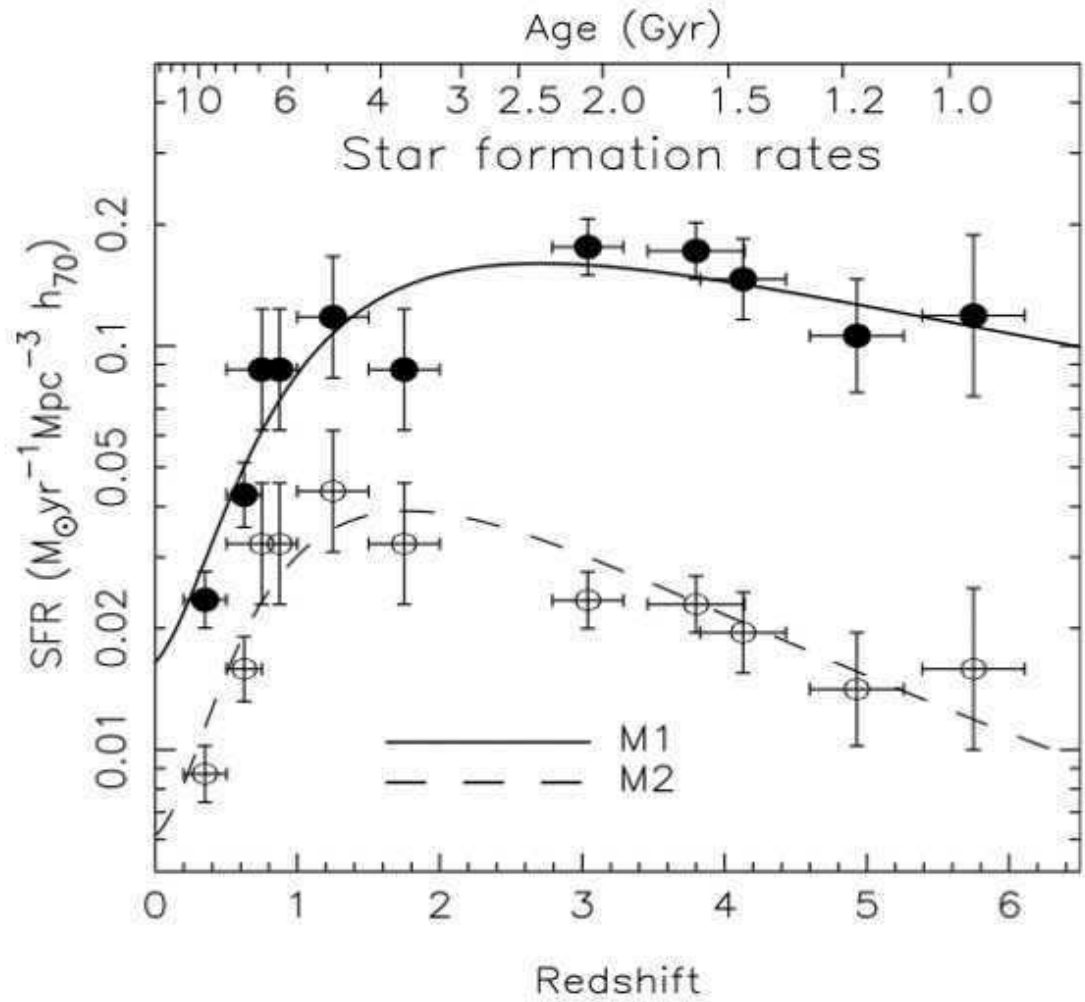}
	\caption{The star formation rate history models (M1 and M2), shown
relative to measurements of the star formation rate at various redshift intervals.}
	\label{fig:sfr}
\end{figure}

We therefore chose to include an analysis for two SFR models. Using a modified
version of the parametric form suggested by~\citet{1998MNRAS.297L..17M}, we
assume SFR$(t)$ evolves as
\begin{equation}
	\text{SFR}(t)=a(t^b e^{-t/c} +d~e^{d(t-t_o)/c}) ,\\
\end{equation}
\noindent where $t$ is given in Gyr. By fitting the measurements of SFR$(z)$ from several
surveys~\citep[see][]{Giavalisco:2003bi}, we determined the coefficients of the
function to be $a=0.182,~b=1.26,~c=1.865,~\text{and}~ d=0.071$ for the
extinction-corrected model (M1), and $a=0.021,~b=2.12,~c=1.69,~\text{and}~
d=0.207$ for the uncorrected model (M2; see Figure~\ref{fig:sfr}). Here $t$ is
the age of the Universe at redshift $z$, and $t_o=13.47$ corresponds to $z = 0$
for both models.

\subsection{The Control Time: The Efficiency of the Survey}\label{sec:controltime}

In comparing predicted yields to what was observed, it is imperative that
corrections are made based on various conditions of the survey. This includes
observational effects such as the magnitude limits, effective sky coverage, and
time over which the survey was conducted, as well as SN type parameters such as
the intrinsic luminosity range, light-curve shapes, and extinction
environments. We combine all of these systematic effects to a single parameter,
the ``control time'' [$t_{c}(z)$], which is in effect the amount of time a SN
at a given redshift could have been observed. We define $t_{c}(z)$ as

\begin{equation}
 	t_{c}(z)=\int_{t}\int_{M_{\lambda}}\int_{A_{\lambda}}P(t\vert M_{\lambda}, A_{\lambda}, z)P(M_{\lambda})P(A_{\lambda})~dA_{\lambda}~dM_{\lambda}~dt,
  	\label{eqn:gentc} \end{equation}

\noindent which is a product of probabilities for observing a SN of specific
absolute magnitude ($M_{\lambda}$, at rest-frame central wavelength $\lambda$)
with specific host-galaxy extinction ($A_{\lambda}$) at specific times ($t$),
summed over all viable absolute magnitudes, extinction values, and time. All
parameters in the equation are dimensionless, except for $dt$ with units of
time.

We determined the sensitivity of our survey in a real-time method, by placing
false SNe of random magnitudes (in the range 23--27 mag) in search-epoch images. 
The modified search images passed through our image-subtraction
pipeline, on to the visual inspection, unbeknownst to the search team. In doing
so, we were able to determine the combination of the intrinsic sensitivity
limits and the search team's efficiency. Only a moderate number ($\sim40$) of false SNe were added to the survey data so that searchers would not become desensitized to real transients by an overwhelming number of bogus detections. 
 To add some realism to the test, the majority of false SNe were added to known galaxies in a Gaussian radial distribution ($0.50'' \pm 0.25''$, $10\pm5$ pixels) truncated at zero radius. It has also been documented that in intermediate and high redshift surveys, a few SNe have been discovered without host galaxies, to within the detection limits of these surveys~\citep{Strolger2002,Gal-Yam2003a,Germany2004a,Germany2004b}. We therefore placed a few SNe in completely random locations, so that searchers would not bias their discoveries based on the requirement of a host galaxy. Astonishingly, 100\% of test SNe were recovered to $m_{F850LP} \le 25$ mag, a testament to the efficacy of the search team and the unparalleled stability of observing conditions with {\it HST}. Beyond
$m_{F850LP} \approx 25$ mag, the recovery efficiency drops rapidly, reaching zero at $m_{F850LP} \ga 26.5$ mag.

The limitation of the of this real-time method was that there were few fake SNe, and therefore only a gross range in detection efficiency could be assessed. This test  cannot appropriately test both rate of decline in efficiency, and the effects of host galaxy light contamination on the efficiency. We therefore independently tested the sensitivity of the survey through a more thorough Monte Carlo simulation.

\begin{figure}
	\centering
 	\includegraphics[angle=-90,width=\textwidth]{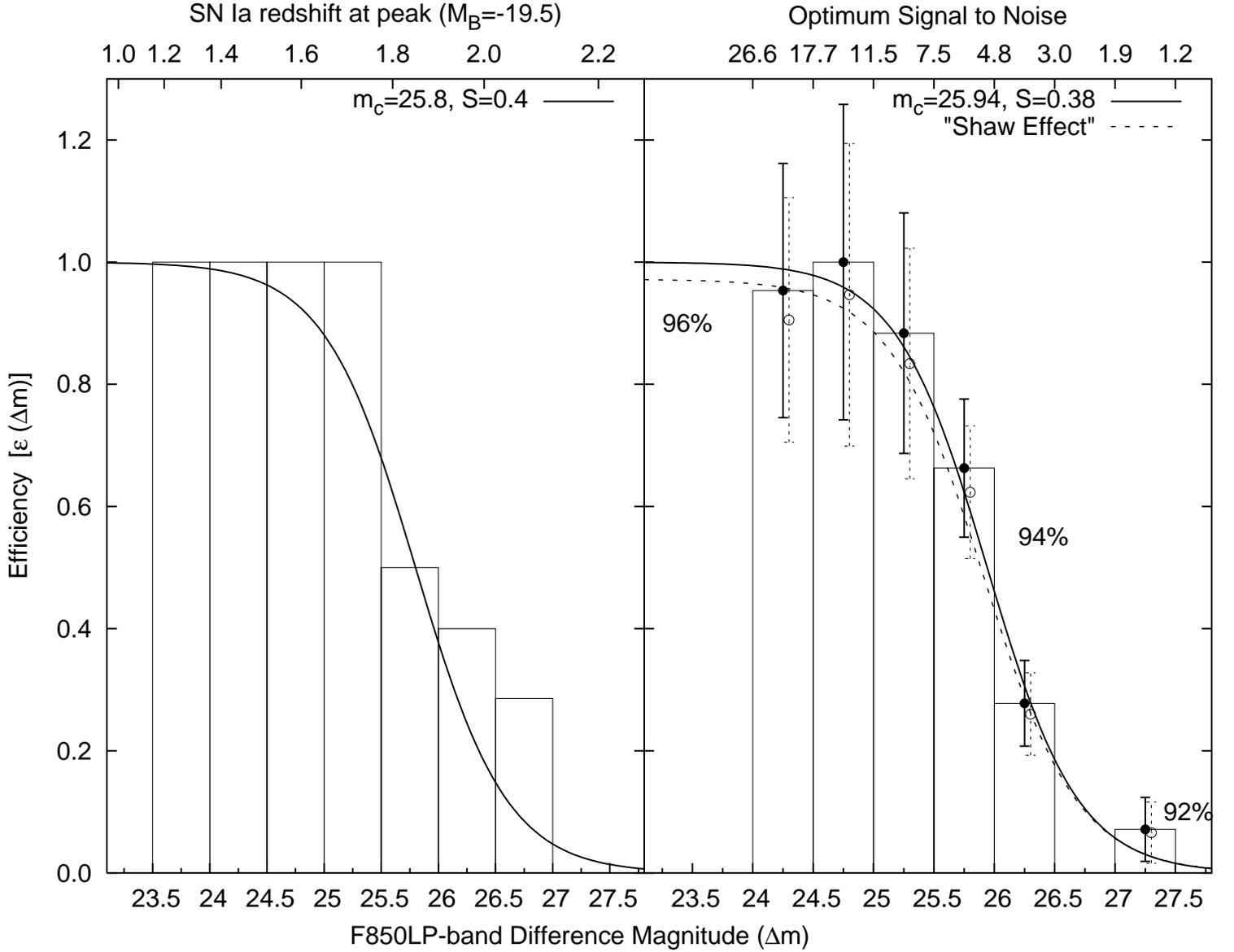}
	\vspace{1.2cm}
        \caption{The efficiency of the survey in recovering false SNe in a pair of
images (discovery and template) of a given difference magnitude. The histograms from both the real-time test (left), and from the Monte Carlo test (right) are shown. The Monte Carlo histogram is shown with Poisson errors (dark circles). The solid line shows the fit of the form $\epsilon(\Delta
m) \propto (1+e^{\Delta m})^{-1}$ used to represent the efficiency. The dotted line (and open circles) show the maximum response to the psudo-Shaw Effect on the efficiency for this survey. For convenience,  we roughly correlate the difference magnitude limits to redshift limits for SNe~Ia discovered near maximum light, and to the optimum signal-to-noise limits for a point source in a residual frame.}
	\label{fig:efficiency}
\end{figure}

500 random host galaxies with phot-$z$ in the range of 1.5 to 2.0 were selected from the GOODS data, and combined to produce light profile of galaxies in this redshift range.  A function was fit to the combined light profile using {\tt galfit}:

\begin{align}
	Bulge&=s_e\times \exp\biggl[-7.688\biggl(\frac{r}{r_e}^{1/4}-1\biggr)\biggr]\\
	Disk&=s_o\times \exp\biggr[- \frac{r}{r_o}+\frac{r_1}{r}^3\biggr]\\
	Total&=Bulge+Disk+Background,
\end{align}

\noindent where the total light profile is well fit by $s_e=0.01,\,r_e=0.055,\,s_o=0.02,\,r_o=3.0,\,r_1=0.0,$ and $Background\approx0$. A set galaxies with phot-$z$ from 1.5 to 2.0 (183 total) were selected in two test tiles, and one fake SN with a random magnitude in the range 25.5--27.5 was added to each of these galaxies with a radial distribution which follows the derived cumulative light profile. A second distribution of faint SNe (24 -- 26.5 mag) were also added to selected bright galaxies with phot-$z < 0.5$ using the same radial distribution as was used for the faint population of galaxies (181 SN total, one per galaxy). These test images were then run through the processing pipeline, and recovered using the automated residual detection algorith. The results of both efficiency tests were combined to produce a histogram of recovered fake SNe as a fraction of the number added (shown in Figure~\ref{fig:efficiency}).

We fit an analytical function to the efficiency distribution following

\begin{equation}     	
	\varepsilon (\Delta m) = \frac{T}{1 + e^{(\Delta m - m_{c})/S}},
	\label{eqn:efficiency} 
\end{equation}

\noindent where $\Delta m$ is the magnitude corresponding to the flux
difference between two consecutive epochs in the $F850LP$ band, determined by

\begin{equation}
 	\begin{split}
     		m_{1}&=ZP - 2.5\times \log(F_{1}) \Rightarrow      F_{1}=10^{-\frac{2}{5}(m_{1} - ZP)}\\
		m_{2}&=ZP - 2.5\times \log(F_{2}) \Rightarrow      F_{2}=10^{-\frac{2}{5}(m_{2}-ZP)}\\
		\Delta m&=ZP - 2.5\times \log(F_{2}-F_{1}).
     	\end{split}
\end{equation}
\\
\noindent Here $ZP$ is the photometric zero point, $T$ is the maximum
efficiency, $m_{c}$ represents a cutoff magnitude where
$\varepsilon(\Delta m)$ drops below 50\% of $T$, and $S$ controls the shape of
the roll-off. As seen in Figure~\ref{fig:efficiency}, the real-time tests show a maximum efficiency
which remains at 100\% ($T=1$) until $\sim25.5$ mag, where background noise begins to play an important role. $\varepsilon(\Delta m)$ drops
with $S=0.4$, reaches the cutoff at $\Delta m_{c}=25.85 \pm 0.1$, and is essentially zero at $\Delta m\ge 27$.  A weighted least-squares fit to the Monte Carlo data show $T=1.03\pm0.09$, $m_c=25.91\pm0.12$, and $S=0.39\pm0.08$. As the maximum efficiency cannot be $>100\%$, we set T=1 as a prior, and found $m_c=25.94\pm0.05$ and $S=0.38\pm0.06$. It is important to note that $m_{c}$ represents only a $\sim 5\sigma$ cutoff. Our simulations show we could detect SNe to within $\sim 3\sigma$, but only a small fraction of the time (depending on the local background light). 

\begin{figure}
	\centering
 	\includegraphics[angle=-90,width=\textwidth]{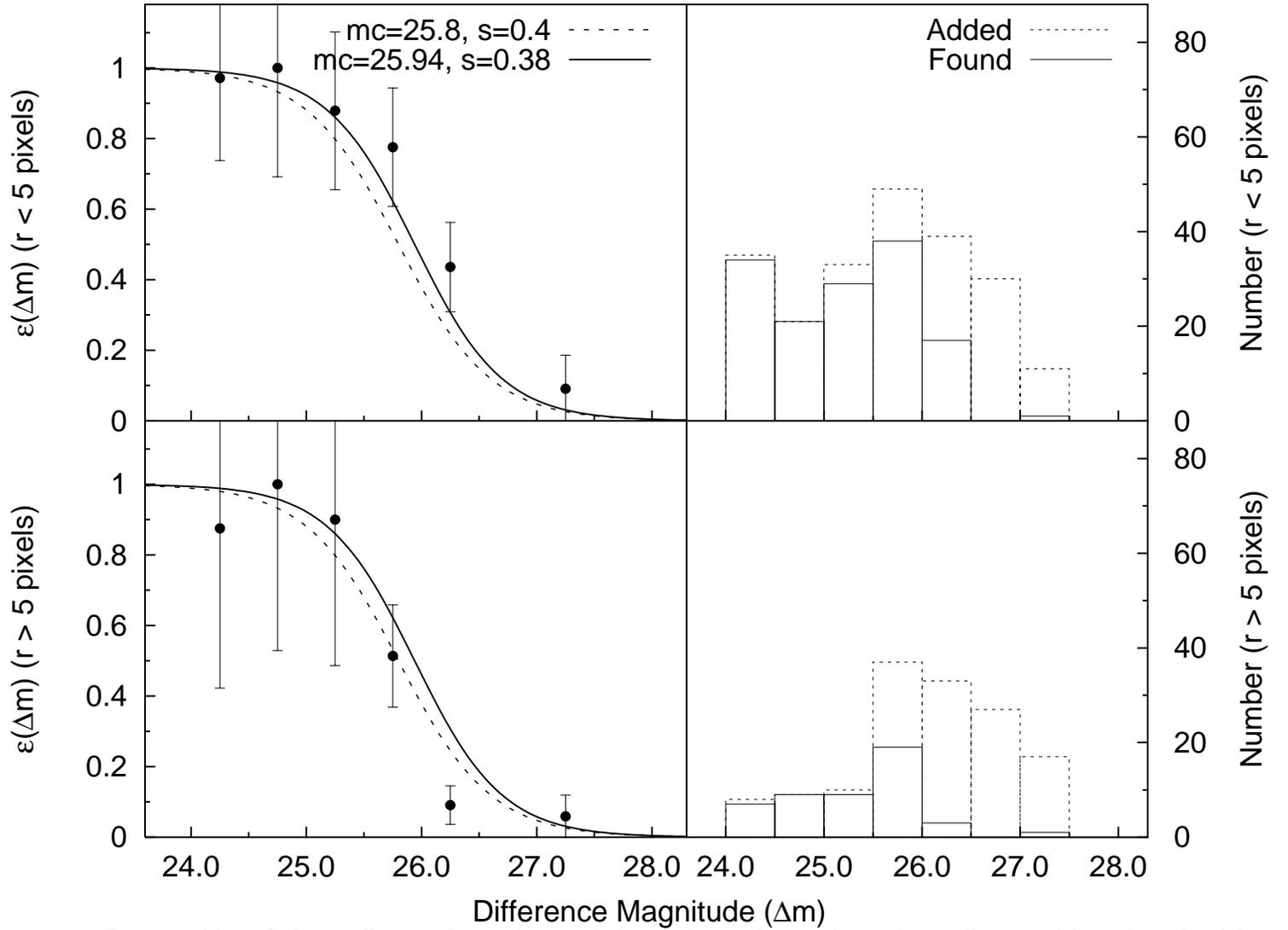}
	\vspace{1.2cm}
	\caption{Monte Carlo test efficiencies for populations which are either nearly coincident with, or well separated from, the nuclei of their host galaxies ($r < 5$ and $r> 5$ pixels, respectively). Neither distribution, specifically the nearly coincident sample, show detectable deviation from fits drawn from the real-time test (dotted line), or the Monte Carlo simulation of the entire sample (solid line).}
	\label{fig:faintradial}
\end{figure}

As with other supernova surveys, it was expected that the efficiency would not only depend on the brightness of the SN, but also the brightness of the host galaxy and the local gradient of light (or synonymously the distance from the host nucleus). Most modern surveys use image subtraction methods to find SNe, and therefore generally do not lose SNe because of overall light contamination from the SN environment, as was the case with the original ``Shaw Effect''~\citep{Shaw1979}. However, faint SNe are lost in the Poisson noise of the host galaxies~\citep[cf.][]{Hardin2000}, or in the residual remaining from an imperfect subtraction of the host galaxy. To account for the possibility of this pseudo-Shaw Effect, we separated the fake SNe into two distributions based on their proximity to the center of the host nuclei, and drew recovery efficiency histograms from the samples (see Figure~\ref{fig:faintradial}). The efficiency histogram drawn from the fake SNe which were nearly coincident with their host nuclei, with radial distances of less than 5 pixels, showed no substantial difference from the histogram drawn from well-separated SNe, indicating that the Shaw Effect was likely insignificant to this survey.  In fact, there was a slight tendency to find more SNe at small radial distances than at larger radial distances. This was an attributed to the automated residual detection algorithm, which also identified the residuals of galaxies due to breathing or focus drift as potential SNe. An important distinction between the real-time method and the Monte Carlo test was that human searchers were capable of distinguishing and rejecting a poor subtraction due to a change in the PSF from a SN candidate, whereas the automated method was not. In reality, regions which showed such PSF residuals were deemed ``unsearchable'' and rejected.

To asses what fraction of  SNe could be lost by rejecting these unsearchable regions (and therefore a potential loss in efficiency), we convolved a test image with a narrow Gaussian filter to produce an image with  PSF $\approx 3$ pixels FWHM, which is 5--10\% larger than the PSFs recorded under the worst conditions of the survey. This convolved image was subtracted from the original image (without the convolution) to produce galaxy residuals which are under-subtracted in the cores. A histogram was drawn from the 2 pixel aperture magnitudes of these residuals, which is well represented by a Gaussian with $\langle\Delta m\rangle=28.42,\,\,\sigma=0.707$. SNe with magnitudes equal to or less than that of a galaxy residual cannot be distinguished from the residual itself, and therefore the flux contained in the narrow region of the core would be unsearchable for SNe of those magnitudes. Accordingly, this flux cannot be included in the total of galaxy light surveyed. As the SN rate is expected to follow the galaxy light, the rejected flux would result in a loss in the overall number of SNe discovered, which we represent as a reduction in efficiency. 

For example, a galaxy core which produced a 22 mag residual would be unsearchable for SNe $>22$ mag. Therefore, the efficiency for SNe $>22$ mag would drop by a fraction proportional to the fraction of all galaxy light that is contained in the cores of galaxies which could produce 22 mag residuals. Fainter galaxy residuals only reduce the efficiency for fainter SNe. These faint galaxy residuals are more numerous, but the flux within the cores of the galaxies which produced them is a considerably smaller fraction of the total flux in the image, and therefore their rejection would result in only a small contribution to the efficiency loss.

We find that only 8\% of the total light in an image was contained in the cores of galaxies, nearly half of which resided in bright galaxies. The galaxies which produced residuals $>23$ mag accounted for approximately $ 4\%$ of the flux in the image, and therefore an overall 4\% efficiency reduction for all SNe. Increasingly fainter galaxies further reduced the efficiency for fainter SNe, leading to a 6\% efficiency drop by $\Delta m=m_c$, and a 8\% reduction by $\Delta m=27.5$. As can be seen in Figure~\ref{fig:efficiency}, the pseudo-Shaw Effect does not significantly reduce the efficiency. As this test involved the worst possible conditions of the survey, it serves only as an upper limit to the impact on the efficiency.

The survey efficiency was used to determine the probability of detecting SNe~Ia
of all redshifts at any given time [$P(t)$ in eq.~\ref{eqn:gentc}]. To do so,
it was important to use a SN~Ia light-curve model that has well observed
multi-wavelength data extending to the rest-frame $U$ band. We used SN~1994D
(R. C. Smith 2003, private communication), a luminous yet
``normal''~\citep[cf.][]{Branch19932383} SN~Ia with $UBVRI$ observed light
curves. Studies have shown that this SN was relatively blue in $U-B$ (by as much as 0.3 mag) compared to normal SNe~Ia at early epochs~\citep{2002PASP..114..833P}.  We therefore attempt to correct for this color excess in the template by applying a linear color correction which is 0.3 mag when the central wavelength of the $F850LP$ filter matches, or is blueward of the rest-frame central wavelength of the $U-$band filter, and gradually decreases to 0.0 mag when the $F850LP-$band matches or is redward of the rest-frame $B-$band. The light curves of SN~1994D were adjusted to the rest frame, and relative to maximum light.

The apparent brightness of a SN~Ia depends on a its luminosity, the age of the SN, the filter in which it is observed, its local host extinction, and the distance of the event. For a SN~Ia of a given absolute magnitude ($M_{peak}$), redshift [or luminosity distance, $D_L(z)$], and extinction ($A_{\lambda}$), we chose intrinsic light curve, $\mathcal{M}_{\lambda}(t'_a)$, of the SN~Ia model in the rest-frame passband which most closely matches the observed $F850LP$ band. The apparent $F850LP$ magnitude at any point in the model light-curve was determined by

\begin{equation}
	\begin{split}
	m&_{F850LP}(t'_a, z)=M_{peak}+\mathcal{M}_{\lambda}[t'_a\times(1+z)^{-1}]+(U-B)_{94D}\\
	&+K^{\lambda}_{F850LP}[z,t'_a\times(1+z)^{-1}]+A_{\lambda}+5~ log(D_{L}(z))+25.
	\label{eqn:appmag}
	\end{split}
\end{equation}

We assumed that SNe~Ia are nearly homogeneous events, with a luminosity at peak of $M_{peak,B}=-19.5\pm\Delta M_{peak}$. The $(U-B)_{94D}$ parameter corrects for the for the U-B color of the template (as described above), and $K^{\lambda}_{F850LP}$ is the K-correction from the rest-frame bandpass to the $F850LP$ band. $t'_a$ is the modified age of a SN~Ia relative to the epoch of maximum light in the $B-$band (see below). We further assumed the intrinsic $B-V$ color of SNe~Ia at peak is 0.0~\citep{lira1995}.

 It has been shown that there is a dispersion in peak absolute magnitudes of SNe~Ia, and that the relative peak luminosity of the events relates to the rate in which their light curves evolve from maximum light. Luminous SN~1991T-like SNe~Ia decline in brightness more slowly than more normal SNe~Ia, and under-luminous SN~1991bg-like SNe~Ia fade more rapidly from peak brightness. Several methods have been developed to account for this relation, e.~g.~the $\Delta M_{15}(B)$ method~\citep{Phillips1993}, the ``stretch'' method~\citep{Perlmutter1997}, and the multicolor light-curve shape algorithm~\citep{RPK199647388}. To account for the heterogeneity of SN~Ia peak luminosity, and the corresponding effect on the light-curve evolution, we used a combination of the most recent adaptation of $\Delta M_{15}(B)$ method~\citep{Phillips1999} and the stretch method~\citep{Perlmutter1997}. The $\Delta M_{15}(B)$ parameter is related to the peak luminosity by the~\citet{Phillips1999} relation,

 \begin{equation}
	\Delta M_{peak}=0.786(\Delta m_{15}(B)-1.1)+0.633(\Delta m_{15}(B)-1.1)^2,\\
\end{equation}

\noindent which is well suited for SNe~Ia in the range of $0.7<\Delta M_{15}(B)<1.7$, extending from the most luminous and slowly declining, to the less luminous, yet normal SNe~Ia. However, it does not appropriately account for the SNe~Ia similar to SN~1991bg, which evolve very rapidly and are intrinsically several magnitudes fainter than SNe~Ia in the normal range. Indeed, the~\citet{Phillips1999} relation is $\sim 1.5$ magnitudes brighter in the $B-$band than has been observed for SN~1991bg-like SNe in the range $1.7<\Delta M_{15}(B)<2.2$. We therefore applied a correction to the relation for SNe~Ia in this range of $\Delta M_{15}(B)$,

\begin{equation}
	\Delta M_{peak}=1.35+0.786(\Delta m_{15}(B)-1.1)+0.633(\Delta m_{15}(B)-1.1)^2.
\end{equation}

\citet{Li2001ApJ546734L} have found that the distribution of SNe~Ia favors normal events, with only $\sim20\%$ of events in the range $0.7 < \Delta m_{15}(B) < 0.9$ (SN~1991T-like SNe), about 20\% in the $1.7 < \Delta m_{15}(B) < 2.2$ range (SN~1991bg-like SNe), and the remaining 60\% in the $0.9< \Delta m_{15}(B) < 1.7$ range. We attempted to characterize this observed distribution by assuming the intrinsic  dispersion in $\Delta M_{15}(B)$ is Gaussian, centered at $\Delta M_{15}(B)=1.1\pm0.35$ and truncated at $\Delta M_{15}(B)<0.7$ and $\Delta M_{15}(B)>2.2$. The implied distribution in peak luminosity is in agreement with the observed distribution from~\citet{2002AJ....123..745R}, and was used as the probability of observing a SN~Ia of a given luminosity [$P(M_{\lambda})\Rightarrow P(\Delta M_{15}(B))$].

A simple way to quantify the effect on the light-curve evolution is by the stretch parameter, which effectively scales the time axis of light curve. \citet{Perlmutter1997} give a relation for the stretch parameter to the  $\Delta M_{15}(B)$ parameter,

\begin{equation}
	stretch=\biggl[\frac{1.96}{\Delta M_{15}(B)-1.1+1.96}\biggr].
\end{equation}

\noindent The modified age of the SN~Ia relative to maximum light was then, $t'_a=t_a\times stretch$, where the actual age, $t_a$, is scaled by the stretch factor. We further adopted $t'_a$ as the epoch in which the first image was taken, and $t'_a+45$~d to be when the second-epoch image was observed. We stepped
through viable values of $t'_a$ (from $\sim 300$~d before to $\sim 200$~d after
maximum light), each time determining $m_{F850LP}(t'_a, z)$, $m_{F850LP}(t'_a+45,
z)$, $\Delta m$, and $\varepsilon (\Delta m,t'_a,z)$. The function $\varepsilon
(\Delta m,t'_a,z)$ serves as a normalized probability function for detecting a SN
at $z$ at time $t'_a$ relative to peak in the observer's frame
[$P(t'_a)\equiv\varepsilon (\Delta m,t'_a,z)$].

The distribution of intrinsic extinction of SNe~Ia due to host galaxies has
been well studied at low redshift. \citet{1999ApJS..125...73J} has shown for 42
SNe (and 4 calibrators) that the extinction distribution is fairly exponential,
with the form $\phi(A_{V}) \propto e^{- A_{V}}$. Assuming the
wavelength-dependent cross-sections of scattering dust to be proportional to
$\lambda^{-1}$, and that $A_{\lambda}\propto\lambda^{-1}$, we adopt,

\begin{equation}     P(A_{\lambda}) \propto e^{- A_{\lambda}}.     \label{eqn:extdistia} \end{equation}

\noindent As both of our survey fields were outside of the Galactic plane ($\vert b\vert > 54\arcdeg$), it was assumed that the Galactic extinction is negligible. 

The total probability for SNe~Ia at redshift $z$ was the sum of the above
probabilities for all viable SN ages. We define the probability as an effective time in
which a SN at $z$ can be detected in our survey by multiplying by the step in
$t'_a$:

\begin{equation}
	\begin{split}
  	t_{c}(z)=&\int_{t'_d}\int_{A_{\lambda}}\int_{\Delta M_{15}(B)} \varepsilon (\Delta m,t'_a,z)\times e^{-(\Delta M_{15}(B)-1.1)^{2} / 0.245}\\&\times e^{-A_{\lambda}/0.347}~d[\Delta M_{15}(B)]~dA_{\lambda}~dt'_a+Const.
	\end{split}
	\label{eqn:tc}
\end {equation}

As a final note on the efficiency of the survey, one might notice from Figure~\ref{fig:fieldhdfn} that the distribution of SNe in the HDFN survey field appears conspicuously asymmetric, possibly indicating an effect (physical or observational) that is unaccounted for in the calculation of the control time. However, the astrometry and redshifts from the GOODS photometric redshift catalog~\citep{Mobasher:2003bm} shows no significant large scale voids or ``pockets'' in regions which have produced few SNe. It is always difficult to determine the significance of an apparent asymmetry after the fact, but we have attempted to do so using Monte Carlo simulations. From randomly placing 23 SNe in an area the size of the HDFN, and bisecting the area in several different ways, we find that asymmetries similar to the observed one can be drawn from random distributions a fair fraction of the time ($>20\%$ K-S probability), although the observed distribution is not the most probable one. Therefore, we treat this apparent asymmetry as a small-number statistical coincidence, and do not make attempts to correct for it. 

\subsection{The Delay Time Models}

\citet{1994MNRAS.268..871T} suggest a general delay time distribution model
that can be represented by an exponential function. This $e$-folding
distribution has been often used
(e.g., ~\citeauthor{1998MNRAS.297L..17M}~\citeyear{1998MNRAS.297L..17M},
\citeauthor{Gal-Yam:2003ed}~\citeyear{Gal-Yam:2003ed}) to explore progenitor
constraints and predict SN rates at high $z$. In this model, it is assumed that
the SNe~Ia are SD systems in which the main-sequence lifetimes of 0.3--3
$M_{\Sol}$ companion stars are chiefly responsible for the delay from formation
to explosion. This model also generally accounts for some additional lagtime to
allow the 3--8 $M_{\Sol}$ progenitor to first become a WD.

Although this model is used more so for its mathematical convenience than for
its physical basis, it is not entirely devoid of the
latter. \citet{1998ApJ...503L.155K} assume two SD scenarios for companion
stars, one involving a red-giant companion with $M_{RG,0} \approx 1 M_{\odot}$,
and one with a main-sequence star with $M_{MS,0} \approx 2 - 3~
M_{\odot}$. Observations of binary systems~\citep{1991A&A...248..485D} show
that the initial mass distribution function for companion stars can be
approximated by $N(M_c)\propto M_c^{0.35}$. Using this information, and the
assumption that delay times are primarily dependent on the companion star's
main-sequence lifetime, one can derive the delay time distribution for this SD
model as being $\Phi(t_d)\propto (t_d/10)^{-0.14}$, where $\langle t_{d,
RG}\rangle \approx 7.01$ Gyr and $\langle t_{d, MS}\rangle \approx 1.46$
Gyr. When considering both SD scenarios, assuming the same mass distribution
function, the mean delay time for both RG~+~MS companions is $\langle t_{d,
RG+MS}\rangle \approx 3.60$ Gyr. This is fairly similar to an $e$-folding
distribution for $\tau \la 3$. More detailed models that involve population
synthesis give broadly similar results~\citep{2000ApJ...528..108Y}.

In this analysis, we assume an $e$-folding delay time distribution of the form

\begin{equation}
	\Phi(t_d, \tau)= \frac{e^{-t_d/\tau}}{\tau},
\end{equation}
\\
\noindent where $\tau$ is the characteristic delay time. We do not attempt to
separate the distribution into constituent parts {\it a priori} (i.e., the
progenitor or companion star lifetimes); rather, we investigate the entire time
lag distribution as a whole. However, it should be noted that models which do
include time for WD development tend to require that this lag time be 
$\sim0.5$ Gyr, not contributing significantly to the overall delay time distribution.

\begin{figure}
	\centering
 	\includegraphics[angle=-90,width=\textwidth]{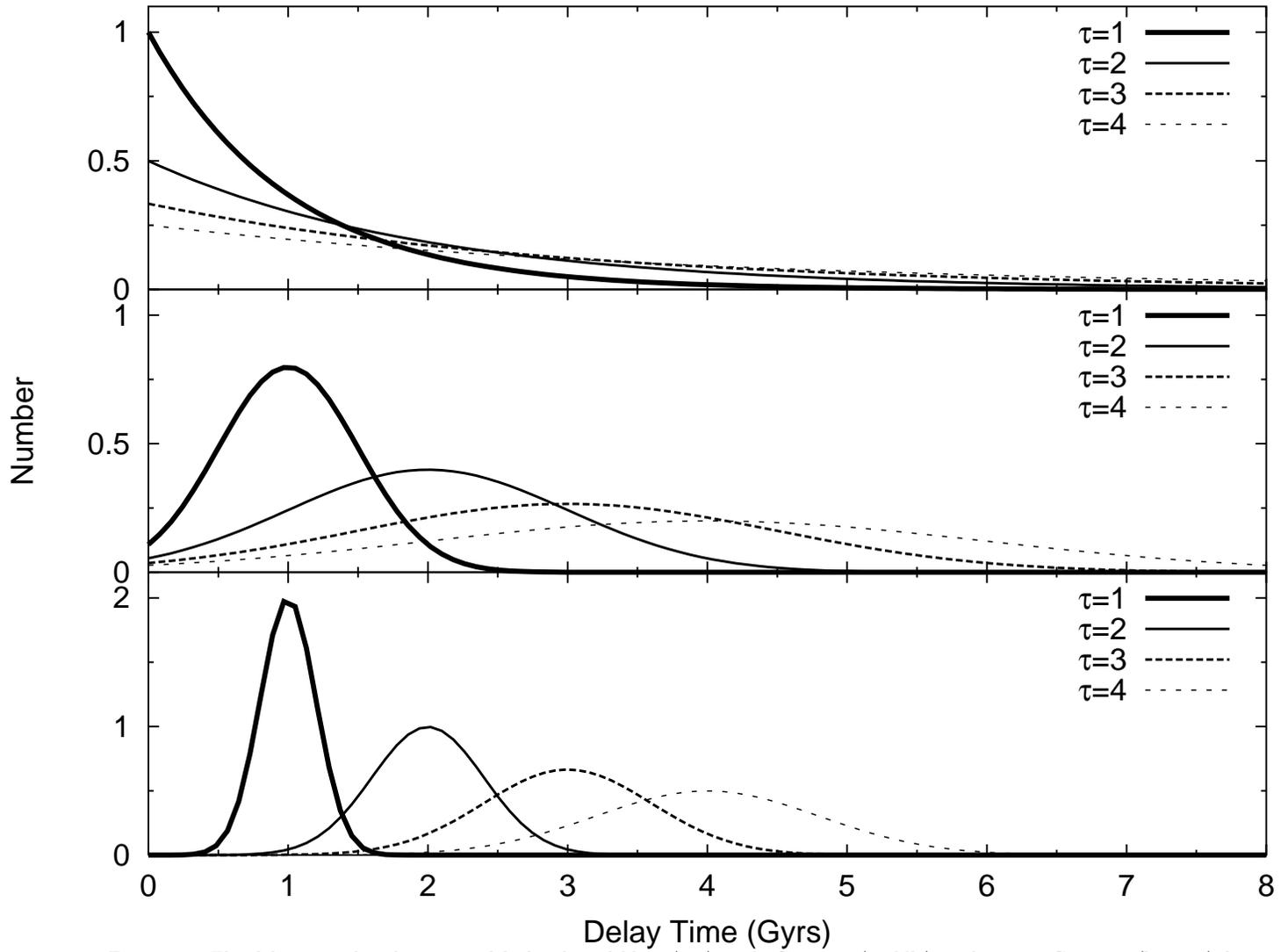}
	\vspace{1.2cm}
	\caption{The delay time distribution models for the $e$-folding (top), 
wide Gaussian (middle), and narrow Gaussian (bottom) functions. Each model is 
plotted with several values of $\tau$.}
	\label{fig:dtmodels}
\end{figure}

Although there is some physical basis in the above $e$-folding model, it is not
reasonable to expect that the delay time distribution is intrinsically
exponential~\citep[see][]{2000ApJ...528..108Y}. It is possible that SNe~Ia
progenitors actually prefer a specific channel to the production events (marked
by a specific delay time) and that there is some scatter in this channel which
leads to a dispersion of delay times, and ultimately a dispersion in SN~Ia
characteristics. An example of using a simple model with a preferred delay time was used by~\citet{Dahlen1999}. To account for this possibility, we chose to further consider Gaussian
functions of two characteristic widths:

\begin{equation}
	\Phi(t_d, \tau)= \frac{1}{\sqrt{2\pi\sigma_{t_d}^2}}e^{-(t_d-\tau)^2/(2\sigma_{t_d}^2)},
\end{equation}
\noindent where our ``wide" and ``narrow" Gaussian models have $\sigma_{t_d} =
0.5\tau$ and $\sigma_{t_d} = 0.2\tau$, respectively. The $\Phi(t_d,\tau)$
models are shown in Figure~\ref{fig:dtmodels} for several values of $\tau$.

\subsection{The Likelihood Test}\label{sec:like}

With assumed SFR$(z)$ and $\Phi(t_d,\tau)$ models, we have used
eqs.~\ref{eqn:numdist} and~\ref{eqn:snriadef} to predict the expected number
distribution of SNe~Ia for the survey. This was compared to the observed
distribution of SNe~Ia to produce a conditional probability test in an
application of Bayes' method:

\begin{equation}
	\begin{split}
	P[\text{Data}\vert \text{SFR}(z),& \Phi(t_d,\tau),\tau]\\&\approx P[\text{SFR}(z), \Phi(t_d,\tau),\tau\vert\text{Data}],
	\label{eqn:probability}
	\end{split}
\end{equation}
\\
\noindent where it was assumed that the SFR$(z)$ model and all other
dependencies (e.g., $\Omega_{M}$, $\Omega_{\Lambda}$, $H_0$, and survey
parameters) are sufficiently well determined that their uncertainties do not
significantly contribute to the overall probability.  The predicted number
distribution, given the assumptions on the models, then served as a probability
function for finding SNe~Ia at the specific redshifts where we have found
them:

\begin{equation}
	\begin{split}
	&P[\text{Data}\vert \text{SFR}(z),\Phi(t_d,\tau),\tau]=\prod_{i=1}^{25} N_{Ia}(z_{i})\\
	 &=\prod_{i=1}^{25}\text{SNR}_{Ia}(z_{i})\times t_{c}(z_{i})\times(1+z_{i})^{-1}\times\frac{\Theta}{4\pi}\times\Delta V(z_i).
	\end{split}
	\label{eqn:sumofprod}
\end{equation}

We normalized the probability distributions to serve as a relative likelihood statistic. Changes in the input model parameters will allow changes in the likelihood with
redshift. Through assuming one of two SFR$(z)$ models (M1 or M2), one of three
$\Phi(t_d,\tau)$ models ($e$-folding, wide Gaussian, or narrow Gaussian), and
several values of $\tau$, we attempted to determine the most likely distribution of
delay times. This will provide important clues to the distribution of channels
for SN~Ia production. 

\begin{figure}
	\centering
 	\includegraphics[angle=-90,width=\textwidth]{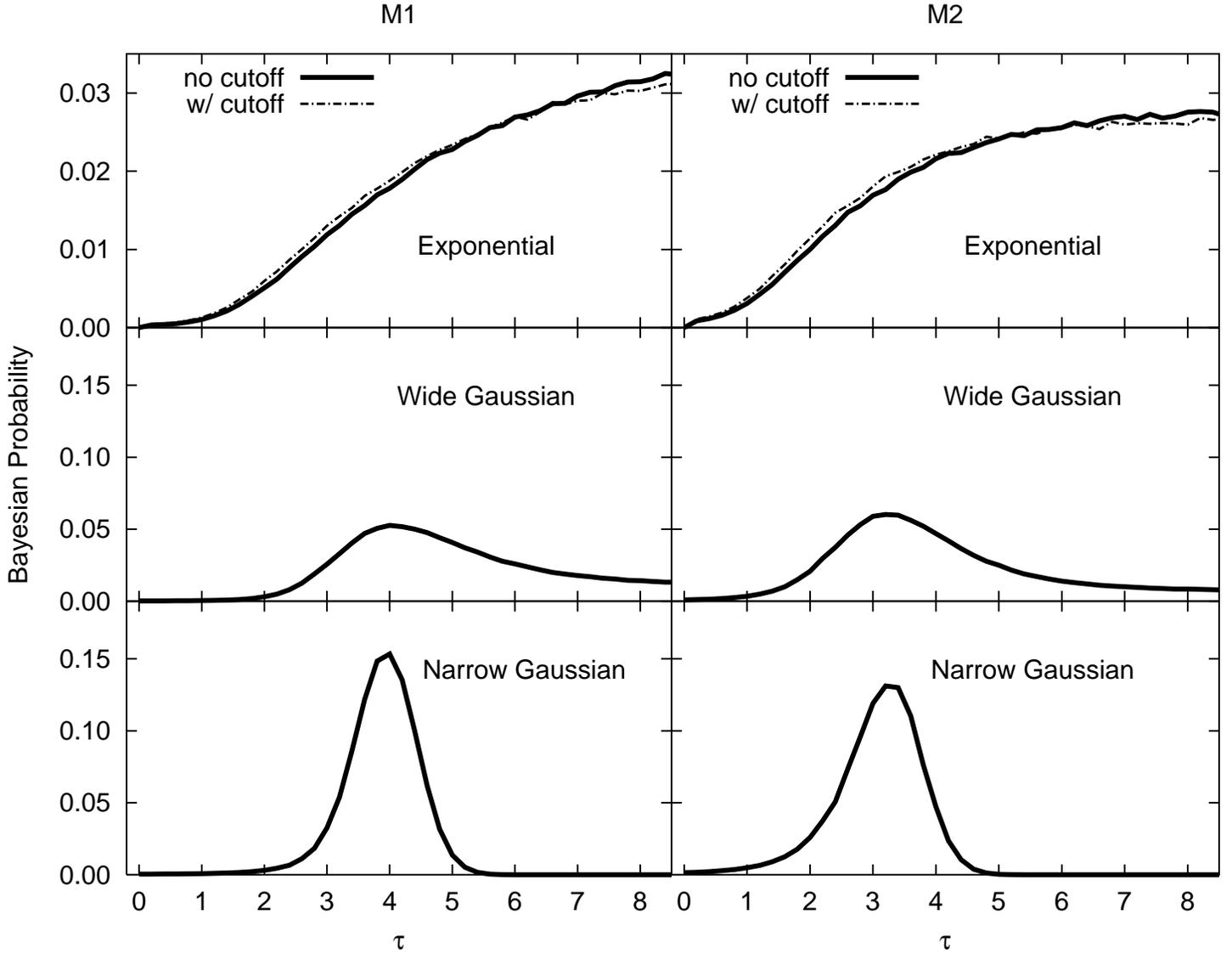}
	\vspace{1.2cm}
	\caption{The probability distributions for the $e$-folding (top panels;
shown with and without metallicity cutoff), wide Gaussian (middle panels), and 
narrow Gaussian (bottom panels) models shown as a function of $\tau$ for the M1 
(left panels) and M2 (right panels) SFR histories. Note that none of the 
$e$-folding models shows a clear maximum likelihood value of $\tau$, and that the 
overall probability values are low for the $e$-folding and wide Gaussian
models. Only in the narrow Gaussian models are there maximum likelihood values 
and overall high probabilities.}
	\label{fig:prob}
\end{figure}

The $P[\text{Data}\vert \text{SFR}(z),\Phi(t_d,\tau),\tau]$ as a function of
$\tau$ is shown in Figure~\ref{fig:prob} for the different $\Phi(t_d)$ and
SFR$(z)$ models. The maximum likelihood $\tau$ values are listed in Table~\ref{tab:stats} for each tested model. The 95\% confidence intervals for each model are also tabulated in Table~\ref{tab:stats}.

\begin{figure}
	\centering
 	\includegraphics[angle=-90,width=\textwidth]{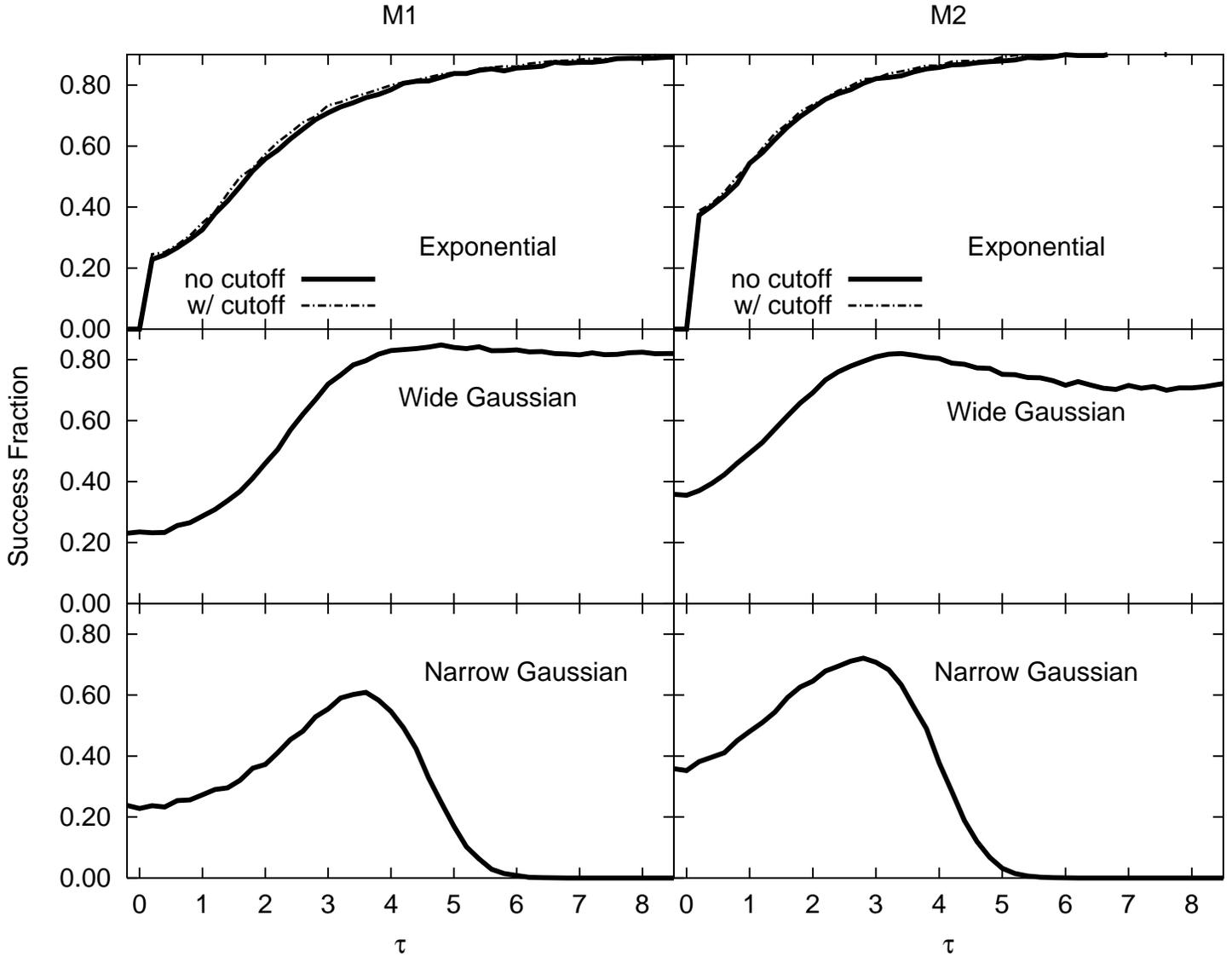}
	\vspace{1.2cm}
	\caption{The fraction of Monte Carlo runs which produce likelihood values equal to or less than the likelihood of the data for each model. Models which could not often produce redshift distributions similar to the observed distribution had a low success fractions (less than $50-60$\%) and were therefore rejected as improbable models for the data. The selected range in success fraction was consistent with the 95\% confidence interval in likelihood for each model.}
	\label{fig:success}
\end{figure}

Although the Bayesian likelihood test gives the most likely values of $\tau$ within a given model, and to some extent, which models are preferred by the data (as the number of free parameters per model are the same), it does not give a very good estimation of which models are inconsistent with the data, and therefore can be rejected at some confidence interval. We attempt to assess how improbable it would be to derive the observed sample from a given model by a Monte Carlo simulation. For each model, an artificial sample of 25 redshifts were drawn from the model distribution 10,000 times, and the likelihood of the test distribution was determined for each run.  We then recorded the success fraction, or the fraction of runs which produced likelihood values less than or equal to the likelihood determined from the observed redshift distribution for the given model. Models which produced redshift distributions similar to the observed distribution less than $50-60\%$ of the time were considered improbable models for the data. The success fractions as a function of $\tau$ for the different $\Phi(t_d)$ and
SFR$(z)$ models are shown in Figure~\ref{fig:success}.

In general, we find that the $50-60\%$ success fraction interval in $\tau$ was consistent with the 95\% confidence intervals determined from the Bayesian likelihood test for each model. However, the selection of the acceptable range in success fraction is somewhat arbitrary. More stringent cuts which seek to either isolate only those models which well reproduce the data, or those which cannot reproduce them at all, will constrict or expand the acceptable range accordingly. We therefore chose to adopt the 95\% interval as our acceptable range in $\tau$ for a given delay time and SFR($z$) model, acknowledging that there may be models in slightly different ranges which could be considered acceptable depending on the selected tolerance level.

\subsection{Results}
 The $e$-folding model showed a preference for large values of $\tau$, with the likelihood of $\tau$ increasing with the value of $\tau$. As the probability distribution remained unbounded at $\tau=10$ Gyr (the limit of our testing region), we chose to consider the 95\% confidence region for $\tau \le 10$ Gyr. This rejected $\tau < 2.6$ and $<2.2$ Gyr to $>95\%$ confidence for M1 and M2, respectively. The trend with increasing $\tau$ can be better exemplified by comparing the $N(z)$ models to the observed $N(z)$. In Figure~\ref{fig:bigplot} the predicted number distribution function of each model for selected values of $\tau$ is compared to the observed $N(z)$, arbitrarily binned with $\Delta z = 0.2$. For
values of $\tau \la 2$, the $e$-folding models require that nearly all SNe~Ia
explode within $\sim 2$ Gyr of progenitor star formation. These ``prompt'' SNe~Ia result in an overestimate of the number of SNe~Ia at $z>1.5$, and do not allow for sufficient development of SNe~Ia at lower redshifts. Increasing the value of $\tau$ increases the fraction of SNe~Ia with delay times over 2 Gyr, and therefore produces higher numbers of lower-$z$ SNe. This alleviates a lot of the skewness in the distribution. However the fraction of prompt SNe~Ia is never less than 10\% of all SNe with delay times below 10 Gyrs, thus the predicted number of SNe~Ia at $z > 1.5$ will always be overestimated for this model. The overall result for large $\tau$ was a distribution which was too wide, overestimating the observed distribution at high and low redshift, and underestimating the vertex of the distribution. We find that this trend existed regardless which SFR model is used. 

An expected behavior of the $e$-folding model is that, at some value of $\tau$, short delay times like $t_d=0.1$ Gyr become as probable as delay times as long as the age of the Universe. The supernova rate then becomes a reflection of the cumulative SFR($z$), and changes little with an increase in $\tau$. This saturation appears to have been reached for $\tau > 7$ Gyr and $>5$ Gyr for the M1 and M2 SFR models, respectively. Therefore, the maximum likelihood values for $\tau$ in the $e$-folding model shown in Table~\ref{tab:stats} are likely a circumstance of the noise in the saturated region. Within our range of modeling, we find that there were only weak maximum likelihood values for $\tau$ in the $e$-folding model using either tested SFR model, and none adequately reproduced the observed frequency distribution with redshift.

The SD progenitor models of \citet{1998ApJ...503L.155K} suggest that a
significant wind emanating from the accreting WD is required to allow a steady
accretion onto the WD, and to extend the range of companion-star
masses. However, in order for this wind to be adequate, the average galactic
metallicity of the Universe must reach [Fe/H] $\ge -1$. In this original
analysis, \citet{1998ApJ...503L.155K} predict this metallicity requirement
imposes a redshift cutoff, beyond which the Universe stops producing SNe~Ia, at
$z\ga1.4$. It is unlikely that this metallicity cutoff could exist at such a
low redshift, as detections of SN~1997ff at $z \approx 1.7$ and SN~2003ak at
$z=1.55$ (from this survey) would be an obvious contradiction. However, in
\citet{2000coex.conf...35N} a refinement was made to allow the distribution of
SNe~Ia to continue to $z \approx 2$, then rapidly decrease in spiral galaxies,
followed by a rapid decrease in elliptical galaxies at $z \approx 2.5$. In this
model, SNe~Ia would not be produced at all beyond $z \ga 3.5$.  To account for
the possibility of a metallicity cutoff (MCO), we executed another test
involving the $e$-folding model where in which the cutoff function as described
by~\citet{2000coex.conf...35N} is applied to the SFR$(z)$. The outcome was very
similar to the results for the $e$-folding model without the MCO, with 95\% confidence intervals of  $\tau > 2.8$ and $>2.0$ Gyr for M1 and M2, respectively. This is not
surprising, considering that the applied MCO would not have a great impact
until $z \approx 2.5$, and the survey was only sensitive to SNe~Ia at $z < 2.0$.

The Gaussian models did, however, show a clear peak in the likelihood functions,
indicating a value of $\tau$ which, for the model, is preferred by the
data. For the wide Gaussian model, the tests show a maximum likelihood at
$\tau=4.0$ Gyr for M1 (3.2 Gyr for M2). Again, as the probability distribution was unbounded at $\tau=10$, we consider $\tau < 2.8$ and $<2.0$ to be rejected with 95\% confidence. Although there appeared to be a statistically preferred value
for $\tau$ for the wide Gaussian model, the predicted distribution shown in Figure~\ref{fig:bigplot} in the range of the best-fit model was still wide, more skewed toward lower redshifts than the observed distribution, and seemingly underestimated the number observed in the $1.2<z<1.4$ range. However, this predicted redshift distribution was much better than the best-fits obtained in the $e$-folding test, which was reflected in the factor of $\sim2$ increase in Bayesian likelihood value.

In contrast to the previously tested models, the width in the range of $t_d$ for the narrow Gaussian model grew weakly with increasing $\tau$, allowing for tests of models without a
significant fraction of prompt SNe~Ia and a much more narrow distribution. The results of our test show a Bayesian maximum likelihood value at $\tau = 4.0$ for M1 (3.2 for M2) which was more than twice as likely as the best-fit model from the wide Gaussian model, and more than four times more likely than the best-fit $e-$folding model. The probability distribution for this model was well bounded by $\tau=10$, indicating that the model was unsupported by the data for large values of $\tau$. We therefore defined the 95\% confidence interval centered on the maximum likelihood value, with  $3.6 < \tau< 4.6$ for M1, and $2.4<\tau<3.8$ for M2. Visually, the predicted $N(z)$ for the narrow Gaussian show a much more convincing match to the observed distributions at the maximum likelihood value than was produced from either of the other tested models, as can be seen in the panel labeled ``BEST FIT'' in Figure~\ref{fig:bigplot}. It appears that the mean (or characteristic) delay time for SNe~Ia can be well constrained and, at least for the narrow Gaussian model, a convincing number distribution with redshift can be drawn.

\section{Conclusions}

Our tests have shown a strong preference by the observed frequency distribution
for delay time distributions in which the majority of SNe~Ia occur more than 2
Gyr from the formation of the progenitor star. All $\Phi(t_d,\tau)$ models
which implied that most SNe~Ia explode within $\sim 2$ Gyr of progenitor
formation show very low likelihoods and are rejected at the 95\%
confidence level. Therefore, SNe~Ia cannot generally be prompt events, nor can
they be expected to closely follow the star formation rate history.

Tests conducted by~\citet{Gal-Yam:2003ed} similarly conclude that the characteristic delay times of SNe~Ia should be large ($>1 - 2$ Gyr) for SFR($z$) models similar to those used in this paper. However, there are a few important differences in these analyses. In~\citet{Dahlen2003} we show from the data presented in this paper that there is a peak in the SN~Ia rate at $z\approx1$.  The data used in~\citet{Gal-Yam:2003ed} study, based on observations from the Supernova Cosmology Project~\citep{pain2002}, do not extend beyond this observed peak, and are limited to $z<0.8$. Moreover, the nature of the $e$-folding $\Phi(t_d,\tau)$ used in~\citet{Gal-Yam:2003ed} is similar to the $e$-fold model used in this paper, except that it accounts for the relatively short main sequence lifetime of the progenitor WD. Therefore, a similar trend is expected in which increasing $\tau$ generally flattens the expected SN redshift distribution. Due to the limited number and range in the observed SN~Ia redshift distribution, the~\citet{Gal-Yam:2003ed} analysis was only moderately sensitive to the slope of the increase in the SN rate. 

The data presented herein not only covers a much larger range in redshift, but they also appear to be unbounded by the volume surveyed at low redshift and the survey efficiency at high redshift; the combination of which would overestimate the observed number by a factor of $\sim 2$ (assuming the SN rate remains constant with time). It is certainly apparent that the observed sample is bounded by something more intrinsic to the SN~Ia rate history, which we interpret as the star formation history convolved with the SN delay time distribution. The analysis presented in this paper is unique because it probes delay time models which allow for a larger variation in the breadth of the redshift distributions without imposing generally unsupported SFR histories.

Our $e$-folding model, comparable to those previously tested in similar analyses,
cannot adequately reproduce the observed redshift distribution of SNe~Ia from
this survey for $\tau \le 2$ Gyr. This would also be true for the delay time distribution function
inferred from the~\citet{1998ApJ...503L.155K} SD model. We find that $\tau$ must be $\ga 2$ Gyr for the $e-$foliding model at a 95\% confidence. We also find that the $e-$folding model itself becomes untestable at $\tau>5 - 7$ Gyrs as predicted redshift distributions are virtually indistinguishable above this limit. Applying a redshift
cuttoff due to metallicity effects based on the \citet{1998ApJ...503L.155K} SD model only weakly
affected the predicted distributions, and produced similar results. The $e-$folding model with large $\tau$ was statistically acceptable by the data, however upon visual comparison with the observed sample, there were apparent inconsistencies with number observed at $z\ge1.5$, and the strength of the vertex of the distribution. The $e-$folding model for large $\tau$ is similar to the DD models shown in~\citet{1994MNRAS.268..871T} and \citet{1998ApJ...497L..57R}, and therefore, these DD models  cannot be significantly rejected.   However the relatively low likelihoods from the Bayesian analysis present here suggests that this mechanism for SN~Ia production is unlikely the dominant channel used by SN~Ia progenitors. We also note that the detection of H$\alpha$ in the spectra of SN~2002ic~\citep{Hamuy03} cannot by itself be taken as evidence against the DD scenario~\citep[see][]{Livio03}.

We tested two Gaussian delay time distribution models. From the maximum
likelihood tests, we find that a narrow dispersion of 1/5 the mean delay time
is significantly more favored than a wide dispersion (1/2 the mean delay). This narrow
Gaussian model also better reproduces the observed redshift distribution of SNe~Ia. 

In Figure~\ref{fig:dndt} we show our best-fit models for the three delay time
distribution functions. \citet{2000ApJ...528..108Y} explore in detail four
evolutionary channels which possibly produce SNe~Ia: the ignition of C in the
core of a merged DD system, the ignition of central C induced by ignition in an
accreted shell (commonly called edge-lit detonation or ELD) from a He-rich RG
companion, ELD induced from a H-rich subgiant or MS companion, and the central
C ignition from normal accretion (no ELD) from a subgiant or MS
companion. Figure 2 of \citet{2000ApJ...528..108Y} shows the expected delay
time distributions for each channel. We reproduce the predicted distributions for the DD and MS models
in Figure~\ref{fig:dndt} of this paper for comparisons to our best-fit models. As can be seen,
there is some similarity between the~\citet{2000ApJ...528..108Y} subgiant
companion models and our best-fit Gaussian models, specifically the narrow
Gaussian model, which is also largely inconsistent with what is expected from
their DD models.  This similarity can also be seen in comparison to general
WD+MS models suggested by \citet{1996ApJ...470L..97H},
\citet{1999ApJ...522..487H}, \citet{1998ApJ...497L..57R}, and
\citet{Han:2003uj}. Our best-fit model does appear similar to the MS+WD
(with ELD) models in the range of the distribution, but it is different
in that the width is larger and the peak is a few Gyr later than what is expected from these
models. This may suggest that these models are largely inconsistent with the data. However,  It should be noted that the testing done in this paper does not
exhaustively cover the possible range in characteristic delay times or widths
of the distribution, and therefore some dissimilarity is expected.

\begin{figure}
	\centering
 	\includegraphics[angle=-90,width=\textwidth]{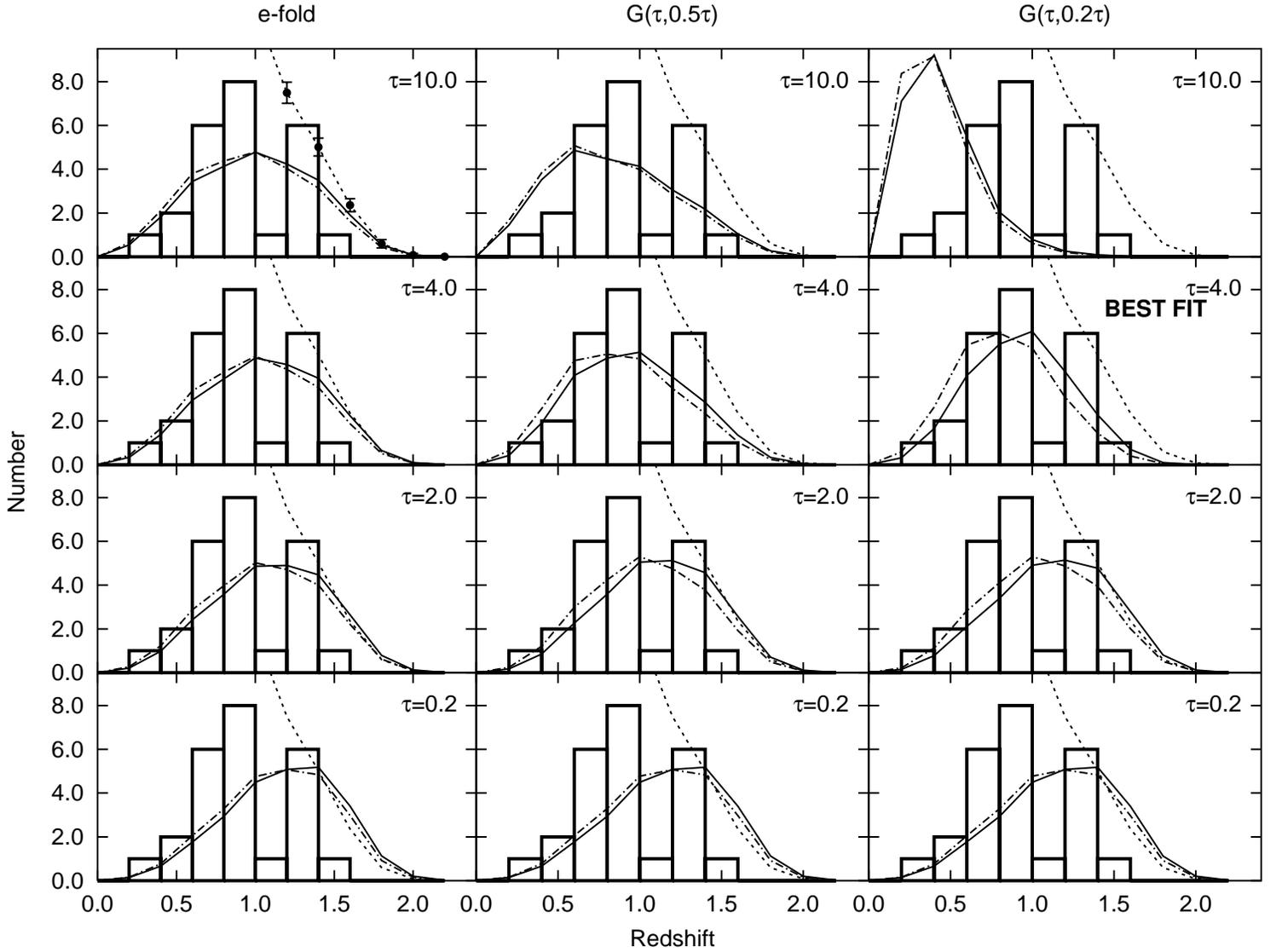}
	\vspace{1.2cm}

	\caption{The predicted number distributions of SNe~Ia for each model
for selected values of $\tau$. The solid line is for the M1 SFR$(z)$ and the
dash-dotted line is for the M2 model. The dotted line shows the control time (or survey efficiency, scaled) with redshift. The systematic effects on the control time are shown in the top left panel (black points).These predicted distributions are compared
to the observed number distribution of SNe~Ia from this survey. Most models cannot adequately reproduce the observed redshift distribution. Only for
the narrow Gaussian model in the range of $\tau \approx 4$ Gyr does the predicted
distribution appear similar to the observed distribution.}
	\label{fig:bigplot}
\end{figure}

\begin{figure}
	\centering
 	\includegraphics[angle=-90,width=\textwidth]{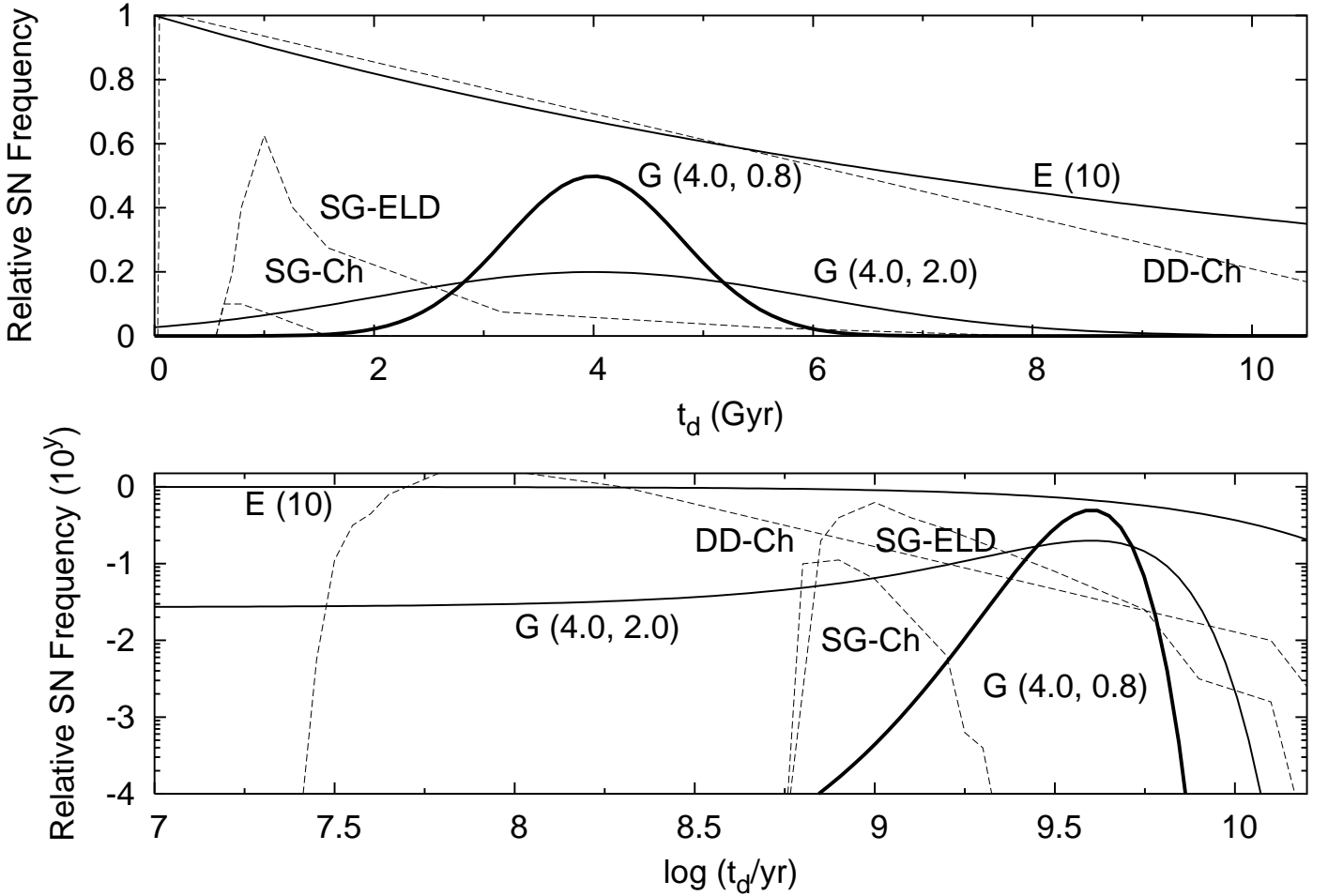}
	\vspace{1.2cm}
	
	\caption{SN distributions, in linear and log space, for the maximum likelihood
values of $\tau$ of each delay time function (solid lines). Shown are the $e$-folding
$\Phi(t_d)$ for $\tau = 10$ [E(10)], the wide Gaussian for $\tau = 
4.0$ [G(4.0, 2.0)], and the narrow Gaussian for $\tau = 4.0$ [G(4.0, 0.8)]. The dashed lines represent predicted distributions of SN~Ia delay times from various
models~\citep[reproduced from Figure 2]{2000ApJ...528..108Y}. Shown are the
predictions from double degenerate mergers (DD-Ch), edge-lit detonations from subgiant donors (SG-ELD),
and normal accretion/detonations from subgiant donors (SG-Ch). There is some
similarity to our best-fit models.}
	\label{fig:dndt}
\end{figure}

It is also important to note that systematic uncertainties have been largely ignored in this analysis. Uncertainty in our models derive from the uncertainties in the derived control times. These errors stem from uncertainties in $\Delta M_{peak}$ (from the coefficients in the $\Delta M_{15}(B)$ and stretch relations), uncertainties in the $\varepsilon (\Delta m)$ parameters, and the pseudo-Shaw Effect between epochs. When combined in quadrature, they result in systematic uncertainties which do not significantly effect the efficiency with redshift for the survey. The systematic uncertainties on the control times are shown in Figure~\ref{fig:bigplot}.  We, therefore, do not account for these errors in the Bayesian analysis.  

This analysis has used all transients identified as SNe~Ia in Table~\ref{tab:sntable}, regardless of the confidence in the identification. However, it is known that some SN~Ib/c can have light curves and colors which are similar to SNe~Ia. Therefore, some SNe~Ia could have been misidentified as Bronze SNe~CC, and conversely some SNe~Ib/c may pollute our Bronze SN~Ia category. However, there were no Bronze SNe~CC at $z>1$, and only one Bronze SN~Ia at $z>1$.  If we considered all Bronze SNe~CC as additional SNe~Ia, the overall number will increase at lower redshifts, but there would be no additional SNe~Ia at $z>1$. Removing all Bronze SNe~Ia also does not greatly affect the high-$z$ sample. Neither rejection would relax the requirement of a substantially large mean delay time, thus the most significant conclusion of this study would remain intact. One could, however, expect minor changes to the width of the best-fit delay time distributions. 

The key implications of our results are that SNe~Ia are not prompt events, and generally require at least $\ge 2$ Gyr to explode from formation. It is also likely that SN~Ia progenitors prefer a specific
channel to explosion, marked by a mean delay time of perhaps as long as $\sim 4$ Gyr, with some scatter in the conditions of the channel. While the implied delay appears to be surprisingly long, this channel is apparently in the range of single degenerate systems which accrete from a main-sequence, or somewhat evolved, non-degenerate companions. The channel would be similar to that which produces supersoft X-ray sources~\citep{1995mpds.conf..105L,Livio99,2003ApJ...588.1003H}.

We thank Dan Maoz for his valuable comments and suggestions which have greatly contributed to this manuscript. Financial support for this work was provided by NASA through programs GO-9352
and GO-9583 from the Space Telescope Science Institute, which is operated by
AURA, Inc., under NASA contract NAS 5-26555. Some of the data presented herein
were obtained at the W. M. Keck Observatory, which is operated as a scientific
partnership among the California Institute of Technology, the University of
California, and NASA; the Observatory was made possible by the generous
financial support of the W. M. Keck Foundation.
The work of D.S. was carried
out at the Jet Propulsion Laboratory, California Institute of Technology, under
a contract with NASA.

\clearpage

\clearpage
\begin{deluxetable}{lcccccccccc}
\tabletypesize{\scriptsize} 	\tablewidth{0pc}
\tablecaption{HHZSS+GOODS Supernovae\label{tab:sntable}}
\tablehead{\colhead{SN}& \colhead{Nickname}& \colhead{U. T.}&
\colhead{R. A. (2000)}& \colhead{Decl. (2000)}& \colhead{SN Type}&
\colhead{Confidence}& \colhead{Redshift}& \colhead{Source}& \colhead{N
(arcsec)}& \colhead{E (arcsec)}}

\startdata

2002fv&Apollo\tablenotemark{a}&2002 Sep
19.6&03:32:22.73&-27:51:09.4&CC&bronze&\nodata&\nodata&\nodata&\nodata\\
2002fw&Aphrodite&2002 Sep
	19.9&03:32:37.52&-27:46:46.6&Ia&gold&1.30&spectrum&0.21&-0.51\\
2002fx&Athena&2002 Sep
	20.8&03:32:06.80&-27:44:34.4&Ia&silver&1.40&spectrum&-0.04&-0.09\\
2002fy&Hades&2002 Sep
	20.9&03:32:18.12&-27:41:55.6&Ia&silver&0.88&phot-z&0.00&0.00\\
2002fz&Artemis&2002 Sep
	21.6&03:32:48.54&-27:54:17.6&CC&silver&0.84&spectrum&-1.66&1.30\\
2002ga&Atlas&2002 Sep
	22.5&03:32:32.62&-27:53:16.7&Ia&bronze&0.99&spectrum&-0.08&0.21\\
2002hp&Thoth&2002 Nov
	01.5&03:32:24.79&-27:46:17.8&Ia&gold&1.30&spectrum&0.02&-0.01\\
2002hq&Re&2002 Nov
	01.5&03:32:29.94&-27:43:47.2&CC&bronze&0.67&spectrum&-0.18&-0.91\\
2002hr&Isis&2002 Nov
	01.6&03:32:22.57&-27:41:52.2&Ia&gold&0.53&spectrum&0.05&0.03\\
2002hs&Bast\tablenotemark{b}&2002 Nov
	02.2&03:32:18.59&-27:48.33.7&CC&bronze&0.39&spectrum&2.50&-0.27\\
2002ht&Osiris&2002 Nov
	02.5&03:32:09.32&-27:41:29.3&Ia&bronze&0.90&phot-z&0.48&0.34\\
2002kb&Denethor&2002 Dec
	20.1&03:32:42.42&-27:50:25.4&CC&gold&0.58&spectrum&-0.20&-0.01\\
2002kc&Bilbo&2002 Dec
	21.5&03:32:34.72&-27:39:58.3&Ia&gold&0.21&spectrum&-0.85&-0.28\\
2002kd&Frodo&2002 Dec
	21.6&03:32:22.34&-27:44.26.9&Ia&gold&0.74&spectrum&-0.98&-3.13\\
2002ke&Smeagol&2002 Dec
	21.9&03:31:58.77&-27:45:00.7&CC&bronze&0.58&spectrum&-0.26&1.15\\
2002kh&Balder&2003 Jan
	04.3&12:36:16.78&+62:14:37.7&Ia&bronze&0.71&phot-z&0.30&-1.20\\
2002ki&Nanna&2003 Jan
	04.6&12:37:28.37&+62:20:39.1&Ia&gold&1.14&spectrum&0.00&-0.10\\
2002kl&Agugux&2003 Feb
	22.0&12:37:49.30&+62:14:06.1&CC&silver&0.41&spectrum&-0.45&0.15\\
2002lg&Prometheus&2002 Jul
	04.2&03:32:35.77&-27:47:58.8&Ia&gold&0.66&spectrum&-0.10&0.13\\
2003aj&Inanna&2003 Feb
	03.2&03:32:44.33&-27:55:06.4&Ia&bronze&1.31&spectrum&-0.08&-0.03\\
2003ak&Gilgamesh&2003 Feb
	03.2&03:32:46.90&-27:54:49.4&Ia&gold&1.55&spectrum&-0.39&0.28\\
2003al&Enki&2003 Feb
	05.7&03:32:05.39&-27:44:29.2&Ia&silver&0.91&phot-z&0.07&-0.03\\
2003az&Torngasak&2003 Feb
	20.9&12:37:19.67&+62:18:37.5&Ia&gold&1.27&spectrum&-0.08&-0.06\\
2003ba&Sedna&2003 Feb
	21.0&12:36:15.88&+62:12:37.7&CC&bronze&0.29&spectrum&0.09&-0.18\\
2003bb&Raven&2003 Feb
	21.6&12:36:24.47&+62:08:35.3&CC&silver&0.95&spectrum&-1.31&0.40\\
2003bc&Michabo&2003 Feb
	21.8&12:36:38.06&+62:09:53.3&CC&silver&0.51&spectrum&-0.55&-1.10\\
2003bd&Anguta\tablenotemark{a}&2003 Feb
	22.0&12:37:25.06&+62:13:17.5&Ia&gold&0.67&spectrum&\nodata&\nodata\\
2003be&Qiqirn&2003 Feb
	22.1&12:36:25.97&+62:06:55.6&Ia&gold&0.64&spectrum&0.00&-0.12\\
2003dx&Phidippides&2003 Apr
	04.5&12:36:31.70&+62:08:48.7&CC&bronze&0.46&phot-z&0.10&0.15\\
2003dy&Borg&2003 Jan
	02.8&12:37:09.14&+62:11:01.2&Ia&gold&1.37&spectrum&-0.35&0.25\\
2003dz&Ashe&2003 Apr
	04.8&12:36:39.91&+62:07:52.7&CC&bronze&0.48&phot-z&0.00&-0.25\\
2003ea&Connors&2003 Apr
	05.7&12:37:12.04&+62:12:38.3&CC&bronze&0.89&phot-z&0.15&0.00\\
2003eb&McEnroe&2003 Apr
	05.7&12:37:15.18&+62:13:34.6&Ia&gold&0.92&spectrum&-0.75&0.50\\
2003en&Odin&2003 Jan
	03.2&12:36:33.12&+62:13:48.1&Ia&bronze&0.54&phot-z&0.10&0.07\\
2003eq&Elvis&2003 May
	24.7&12:37:48.34&+62:13:35.3&Ia&gold&0.85&spectrum&0.10&-0.42\\
2003er&Janice&2003 May
	25.4&12:36:32.27&+62:07:35.2&CC&silver&0.63&phot-z&0.70&-0.70\\
2003es&Ramone&2003 May
	25.5&12:36:55.39&+62:13:11.9&Ia&gold&0.97&spectrum&0.30&-0.49\\
2003et&Jimi&2003 May
	25.7&12:35:55.87&+62:13:32.8&CC&silver&0.83&phot-z&0.14&-0.50\\
2003eu&Lennon&2003 May
	25.7&12:36:05.90&+62:11:01.6&Ia&silver&0.76&phot-z&0.30&-0.70\\
2003ew&Jagger&2003 May
	21.8&12:36:27.78&+62:11:25.1&CC&bronze&0.66&phot-z&-0.10&-0.21\\
2003N&Loki&2003 Apr
	04.7&12:37:09.14&+62:11:01.2&CC&bronze&0.43&spectrum&0.20&0.00\\
2003lv&Vilas&2003 Apr 
        04.7&12:37:28.89&+62:11:28.7&Ia&silver&0.86&spectrum&0.00&0.00\\
\enddata 
\tablecomments{Offsets are given from the center of the host galaxy to
supernova.}
\tablenotetext{a}{No host galaxies were detected for SNe~2002fv and 2003bd to
within the magnitude limits of the survey.}
\tablenotetext{b}{SN~2002hs has at least two neighboring galaxies, the closest
of which had a phot-$z$ = 1.1, and the other had a spectroscopically measured
$z = 0.39$. Light-curve fits to the photometry showed that it was less
consistent with any SN type at $z \approx 1.1$ and more consistent with a
SN~Ib/c at $z = 0.39$.}
	
\end{deluxetable}

\clearpage

\begin{deluxetable}{lccr}
	\tablewidth{0pc}
	\tablecaption{{\it HST} Photometry\label{tab:snphot}}
	\tablehead{\colhead{SN}& \colhead{Filter}& \colhead{JD + 2,452,000}& \colhead{Magnitude}}
	\startdata
2002fv&$F850LP$&536.74&25.07 (0.06)\\&&580.32&25.72 (0.11)\\&&628.29&26.27 (0.18)\\&$F775W$&536.74&25.34 (0.06)\\&$F606W$&536.83&26.54 (0.06)\\\hline2002fy&$F850LP$&490.30&23.77 (0.02)\\&&538.20&25.23 (0.07)\\&&579.72&26.40 (0.20)\\&$F775W$&490.32&24.37 (0.03)\\&&538.19&26.06 (0.12)\\&$F606W$&490.30&26.15 (0.04)\\&&538.18&28.16 (0.25)\\\hline2002fz&$F850LP$&538.96&24.94 (0.06)\\&&578.71&26.30 (0.19)\\&$F775W$&538.92&25.20 (0.05)\\&&578.45&26.73 (0.21)\\\hline2002ga&$F850LP$&488.72&25.90 (0.13)\\&&539.65&24.50 (0.04)\\&&578.95&26.18 (0.17)\\&&628.70&26.79 (0.29)\\&$F775W$&488.70&27.50 (0.41)\\&&539.65&25.26 (0.06)\\&&628.70&28.90 (1.25)\\&$F606W$&488.64&27.87 (0.20)\\&&539.65&26.42 (0.05)\\&&628.70&29.0 (0.53)\\\hline2002hq&$F850LP$&580.90&25.34 (0.08)\\&$F775W$&579.50&26.43 (0.11)\\\hline2002hs&$F850LP$&580.52&25.27 (0.07)\\&$F775W$&580.37&26.05 (0.12)\\&$F606W$&580.32&26.77 (0.07)\\\hline2002ht&$F850LP$&581.00&25.64 (0.10)\\&&630.17&26.13 (0.16)\\&$F775W$&580.95&26.33 (0.15)\\\hline2002kb&$F850LP$&488.72&24.62 (0.04)\\&&539.79&25.10 (0.06)\\&&578.95&25.60 (0.10)\\&$F775W$&488.70&24.83 (0.04)\\&&539.78&25.17 (0.05)\\&&578.80&25.84 (0.09)\\&$F606W$&488.65&25.02 (0.02)\\&&539.71&27.01 (0.06)\\\hline2002ke&$F850LP$&630.11&25.52 (0.09)\\&$F775W$&630.03&26.00 (0.11)\\&$F606W$&630.02&27.60 (0.15)\\\hline2002kh&$F850LP$&642.27&24.62 (0.04)\\&$F775W$&642.27&25.64 (0.11)\\&$F606W$&642.22&28.4 (0.8)\\\hline2002kl&$F850LP$&600.04&24.95 (0.08)\\&&642.66&25.42 (0.09)\\&$F775W$&600.04&25.39 (0.06)\\&&642.66&26.32 (0.14)\\&$F606W$&599.99&26.37 (0.05)\\&&642.57&27.81 (0.18)\\\hline2002lg&$F850LP$&464.83&24.31 (0.04)\\&&489.17&25.89 (0.13)\\&$F775W$&489.19&26.80 (0.22)\\\hline2003aj&$F850LP$&666.3&25.60 (0.10)\\&&673.31&25.40 (0.10)\\&&688.3&26.50 (0.23)\\&$F775W$&673.30&26.60 (0.19)\\\hline2003al&$F850LP$&676.02&24.59 (0.04)\\&$F775W$&676.10&25.56 (0.08)\\&$F606W$&675.89&28.95 (0.51)\\\hline2003ba&$F850LP$&691.24&23.63 (0.02)\\&$F775W$&691.24&24.04 (0.02)\\&$F606W$&691.16&25.03 (0.02)\\\hline2003bb&$F850LP$&692.11&25.59 (0.10)\\&$F775W$&692.10&25.34 (0.06)\\&&732.84&26.17 (0.13)\\&$F606W$&692.45&26.30 (0.05)\\\hline2003bc&$F850LP$&692.35&24.08 (0.03)\\&&733.84&24.66 (0.04)\\&&780.75&25.72 (0.11)\\&$F775W$&692.12&24.10 (0.02)\\&&733.74&25.32 (0.06)\\&&781.06&25.96 (0.11)\\&$F606W$&691.1&25.00 (0.02)\\&&733.84&27.33 (0.12)\\&&781.06&28.40 (0.31)\\\hline2003dx&$F850LP$&732.85&25.27 (0.07)\\
&$F775W$&732.84&25.76 (0.11)\\
&$F606W$&732.81&28.37 (0.31)\\\hline2003dz&$F850LP$&733.86&25.14 (0.07)\\
&$F775W$&733.86&25.39 (0.06)\\
&$F606W$&733.86&26.14 (0.04)\\\hline2003ea&$F850LP$&783.72&25.40 (0.09)\\&$F775W$&783.66&26.00 (0.11)\\&$F606W$&783.65&27.50 (0.14)\\\hline2003en&$F850LP$&642.31&25.60 (0.10)\\&$F775W$&642.30&26.10 (0.20)\\&$F606W$&642.28&27.30 (0.12)\\\hline2003er&$F850LP$&784.51&23.22 (0.02)\\&$F775W$&784.44&23.11 (0.01)\\&$F606W$&784.61&23.21 (0.01)\\\hline2003et&$F850LP$&784.94&25.38 (0.08)\\&$F775W$&784.83&25.54 (0.07)\\&$F606W$&784.92&25.90 (0.04)\\\hline2003eu&$F850LP$&784.94&24.40 (0.04)\\&$F775W$&784.93&25.04 (0.05)\\&$F606W$&784.92&26.79 (0.08)\\\hline2003ew&$F850LP$&646.86&24.21 (0.03)\\&&690.57&24.36 (0.04)\\&$F775W$&646.85&24.83 (0.04)\\&&690.49&25.05 (0.05)\\&$F606W$&646.85&26.92 (0.08)\\&&690.49&27.51 (0.14)\\\hline2003N&$F850LP$&642.21&26.21 (0.18)\\
&$F775W$&642.19&26.43 (0.16)\\
&$F606W$&642.23&27.11 (0.10)\\\hline
	\enddata
	\tablecomments{Magnitudes are in the Vega-based system. Photometric errors are in parentheses. } 
\end{deluxetable}

\begin{deluxetable}{cccccc}	
	\tabletypesize{\small} 	\tablewidth{0pc}	\tablecaption{Likelihood Statistics\label{tab:stats}}
	\tablehead{\colhead{Statistic}&\colhead{SFR Model}&\colhead{$e$-folding}&\colhead{$e$-folding w/ MCO}&\colhead{$G(\tau,0.5\tau$)}&\colhead{$G(\tau,0.2\tau$)}}

	\startdata
	Max. Likelihood $\tau$&M1&$9.8$&$9.8$&4.0&4.0\\
	&M2&$9.8$&$8.2$&3.2&3.2\\
	95\% Interval $\tau$&M1&$>2.6$&$>2.8$&$>2.8$&$3.6-4.6$\\
	&M2&$>2.2$&$>2.0$&$>2.0$&$2.4-3.8$\\
	\enddata
	\tablecomments{95\% interval for narrow Gaussian models are determined symmetrically about maximum likelihood value. All others are given for $\tau > 95\%$ confidence interval. Values are given in Gyrs.}
\end{deluxetable}

 \end{document}